\documentclass[lettersize,journal]{IEEEtran}
\IEEEoverridecommandlockouts
\usepackage{amsmath,amssymb,amsfonts}
\usepackage[ruled,linesnumbered]{algorithm2e}
\usepackage{xcolor}
\usepackage{array}
\usepackage{subfigure}
\usepackage{color} 
\usepackage{setspace}
\usepackage{textcomp}
\usepackage{stfloats}
\usepackage{url}
\usepackage{verbatim}
\usepackage{graphicx}
\usepackage[noadjust]{cite}

\newtheorem{lemma}{Lemma}
\newtheorem{Remark}{Remark}
\def\BibTeX{{\rm B\kern-.05em{\sc i\kern-.025em b}\kern-.08em
		T\kern-.1667em\lower.7ex\hbox{E}\kern-.125emX}}
\begin{document}
	
	\title{Energy Efficiency Maximization for \\Movable Antenna-Enhanced MIMO \\Downlink System Based on S-CSI}
\author{\IEEEauthorblockN{Xintai Chen, Biqian Feng, Yongpeng Wu, \emph{Senior Member, IEEE}, Xiang-Gen Xia, \emph{Fellow, IEEE}, \\and Chengshan Xiao, \emph{Fellow, IEEE}}
	\thanks{X. Chen, B. Feng, and Y. Wu are with the Department of Electronic Engineering, Shanghai Jiao Tong University, Minhang 200240, China (e-mail: chenxintai@sjtu.edu.cn; fengbiqian@sjtu.edu.cn; yongpeng.wu@sjtu.edu.cn).}
	\thanks{Xiang-Gen Xia is with the Department of Electrical and Computer Engineering, University of Delaware, Newark, DE 19716 USA (e-mail: xxia@ee.udel.edu).}
	\thanks{Chengshan Xiao is with the Department of Electrical and Computer Engineering, Lehigh University, Bethlehem, PA 18015 USA (e-mail: xiaoc@lehigh.edu).}
	\thanks{Corresponding author: Yongpeng Wu.}}

\maketitle

	\begin{abstract}
	This paper presents an innovative movable antenna (MA)-enhanced multi-user multiple-input multiple-output (MIMO) downlink system. We aim to maximize the energy efficiency (EE) under statistical channel state information (S-CSI) through a joint optimization of the precoding matrix and the antenna position vectors (APVs). To solve the resulting stochastic problem, we first resort to deterministic equivalent (DE) tecnology to formulate the deterministic minorizing function of the system EE and the deterministic function of each user terminal (UT)'s average achievable rate w.r.t. the transmit variables (i.e., the precoding matrix and the transmit APV) and the corresponding receive APV, respectively. Then, we propose an alternating optimization (AO) algorithm to alternatively optimize the transmit variables and the receive APVs to maximize the formulated deterministic objective functions, respectively. Finally, the above AO algorithm is tailored for the single-user scenario. Our numerical results reveal that, the proposed MA-enhanced system can significantly improve the system EE compared to several benchmark schemes based on the S-CSI and the optimal performance can be achieved with a finite size of movement regions for MAs.
\end{abstract}

\section{Introduction}
In recent years, there has been a significant surge in the demand for wireless data services, driven by the proliferation of mobile devices and the advent of cutting-edge applications such as virtual reality, cloud-based services, and edge computing \cite{Tswi}. Consequently, they present novel challenges for the evolution of future wireless communication systems. Advanced movable antenna (MA, also known as fluid antenna \cite{BLIF})-enhanced multiple-input multiple-output (MIMO) technique proposed lately is a promising solution to tackle these issues. Different from the fixed-position antennas in conventional MIMO systems, it can fully exploit the spatial variations of the wireless channel across the entire transmit/receive region by dynamically reconfiguring the physical placement of antennas \cite{Mapa}. In a few initial works, antenna position vector (APV) optimization is one of the main topics, which aims at the optimal performance of MA-enhanced MIMO systems by using instantaneous channel state information (I-CSI) \cite{Mccf,MAAM,MAET,MAAE,FBCO,SBSE,MEMC,Mapo,Jbaa} or statistical CSI (S-CSI) \cite{JBaA,6MAB,6MAE,CMfB,JBaAEE}.

To explore its potential, the achievable rate serves as an important performance metric for MA-enhanced systems in \cite{Mccf,JBaA,MAAM,MAET}. In \cite{Mccf}, the capacity of the single-user MIMO system is maximized by an efficient alternating optimization (AO) method, where the transmit and receive APVs as well as the covariance matrix of the transmit signal are iteratively optimized. On the other hand, a constrained stochastic successive convex approximation (CSSCA) algorithm is proposed in \cite{JBaA} to maximize the average achievable rate of the single-user MIMO system exploiting S-CSI by jointly optimizing the APVs and the transmit covariance matrix. In \cite{MAAM}, the discrete positions of the MAs are optimized through a greedy search-based algorithm, aiming to enhance the achievable rate of a two-user multicast multiple-input single-input system. Moreover, a low-complexity branch-and-bound-based method is proposed to achieve the optimal multicast rate when the two users experience the same line-of-sight (LOS) path losses. In \cite{MAET}, authors reveal that, to maximize the minimum rate of a two-user multicasting system, the jointly optimal transmit beamforming and APV at base station (BS) can be achieved via separate optimization of the two variables. In particular, the optimal APV is firstly determined via the successive convex approximation (SCA) technique, based on the principle of maximizing the correlation between the channels from BS to the users. Subsequently, the optimal closed-form transmit beamforming is derived through straightforward reasoning. 

Furthermore, the studies have also delved into enhanced multi-beamforming techniques using a linear array of MAs. For example, as detailed in \cite{MAAE}, the optimal solutions for the APV and the antenna weight vector are obtained in closed form, ensuring the maximum array gain in the desired direction while simultaneously facilitating null steering across all undesired directions. Also, in \cite{FBCO}, to maximize the minimum beam gain across the desired spatial regions, the weights and APV are jointly optimized by an AO algorithm. Besides, in \cite{SBSE}, the MAs' positions are adjusted to maximize the minimum beamforming gain across the entire wideband spectrum under the near-field channel model. This optimization is achieved by employing a smoothed gradient-descent-ascent method. 

Additionally, six-dimensional MA (6DMA) system is initially introduced in \cite{6MAB,6MAE,6ewn}, where the three-dimensional positions and three-dimensional rotations of distributed antenna surfaces can be adjusted based on the users’ spatial distribution and S-CSI. In \cite{6MAB}, the channel model between the 6DMA-enabled BS and the user in terms of six-dimensional positions and rotations of antenna surfaces as well as the practical constraints on six-dimensional antennas’ movement are first introduced. Then, an efficient AO algorithm is proposed for maximizing the average achievable rate by leveraging the Monte Carlo (MC) simulation technique. Specifically, each dimension of six-dimensional positions and rotations of each 6DMA surface is iteratively optimized through the conditional gradient method while fixing other variables. Furthermore, the discrete movement of 6DMA at the BS is considered in \cite{6MAE}. The offline MC and online conditional sample mean algorithms are developed for the practical cases with and without statistical channel knowledge of the users, respectively. In \cite{CMfB}, authors consider the practical application of 6DMA in the existing BSs, where the conventional fixed-position uniform planar arrays (UPAs) and 6DMA surfaces are deployed at the BS in a hybrid manner.

Despite various research efforts, deploying MAs with I-CSI in practical systems remains challenging, due to the associated high delays and power consumption. Indeed, these delays arise from various processes such as CSI estimation and/or feedback and advanced signal processing for APVs design as  well as physical antenna movement. The I-CSI can quickly become outdated if these delays exceed the channel coherence time. Besides, power consumption becomes unaffordable in fast fading channel due to frequent movement of the MAs, especially when considering power-saving requirement. In contrast, the S-CSI, e.g., the spatial correlation and channel mean, tends to remain invariant over a longer period. Therefore, leveraging S-CSI for the design of APVs serves as a practical solution for the deployment of effective MA systems.

On the other hand, the traditional spectral efficiency-orient wireless transmission prioritizes data rates over energy efficiency (EE) due to the limited radio spectrum. Nevertheless, the rapid proliferation of data requirements has also led to a substantial increase in power consumption, making EE a critical consideration.  This shift is particularly important as the number of connected devices continues to grow exponentially, underscoring the necessity for more sustainable and efficient communication solutions \cite{Ndfm}. Numerous research have been devoted to the design of EE-orient wireless transmission with fixed-position antennas \cite{Eemm, Eeiw, Eetd, EEOf}. These research typically aims to maximize the ratio of the achievable rate to the corresponding power consumption, reflecting a more comprehensive perspective on system performance \cite{Eeiw, Eetd, EEOf}.
Despite its importance, the system EE of the MA-enhanced system with S-CSI has been evaluated solely in \cite{JBaAEE}, highlighting a critical research gap. In \cite{JBaAEE}, a deterministic upper bound of the EE of a point-to-point system is established by leveraging Jensen's inequality and is maximized by the AO method. However, conducting maximization on an upper bound might potentially lead to a degradation of the optimization effect.

Motivated by the aforementioned discussion, we aim to maximize EE for a power-saving MIMO system that exploits S-CSI in this paper. The main contributions of this paper can be summarized as follows:
	\begin{itemize}
	\item [1)]
	Firstly, we investigate the system EE of MA-enhanced multi-user downlink MIMO systems exploiting S-CSI. To this end, the deterministic equivalent (DE) technique is leveraged to formulate the deterministic minorizing function of the system EE and the deterministic function of each user terminal (UT)'s average achievable rate for the multi-user scenario w.r.t. the transmit variables (i.e., the transmit APV and precoding matrices) and the corresponding receive APV, respectively. This avoids the slow convergence of the CSSCA algorithm \cite{JBaA} and the averaging over massive high-dimensional random states of the MC technique \cite{6MAB}.
	\item [2)]
	Secondly, we develop an AO algorithm to maximize the system EE by iteratively optimizing the transmit variables and the receive APVs based on the rule of maximizing the deterministic minorizing function of the system EE and the deterministic functions of the average achievable rates, respectively. In particular, the deterministic minorizing function is first parameterized with its optimal precoding matrices and then its gradient w.r.t. the transmit APV can be calculated efficiently. Consequently, in each subproblem, we efficiently develop a low-complexity solution with the SCA method.
	\item [3)]
	Thirdly, we tailor the AO algorithm for the single-user scenario. One difference is that, since there is no multi-user interference, the optimal precoding matrix for maximizing the deterministic function of the system EE can be achieved efficiently when the APVs are fixed. Then, the transmit APV can be directly optimized to maximize the deterministic function with the optimal precoding matrix. 
	\item [4)]
	Finally, we provide extensive numerical results to reveal the significant gains in the system EE realized with the proposed MA-enhanced MIMO system compared to several benchmarks with S-CSI, including those employing conventional fixed-position UPAs at BS and/or UTs as well as MAs with discrete movements.
\end{itemize}

The remainder of this paper is organized as follows. Section II presents the system model and problem formulation. Section III provides the proposed AO algorithm. Numerical results and corresponding discussions are presented in Section IV. Finally, Section V concludes this paper.

\textbf{Notations}: Vectors (lower case) and matrices (upper case) are presented in boldface. $(\cdot)^{T}$ and $(\cdot)^{H}$ denote the transpose and conjugate transpose (Hermitian), respectively. The ensemble expectation, matrix trace, and determinant operations are
denoted by $\mathbb{E}\{\cdot\}$, $\operatorname{tr}(\cdot)$, and $\operatorname{det}(\cdot)$, respectively. $\mathbf{A} \succeq \mathbf{0}$ indicates that $\mathbf{A}$ is a positive semi-definite matrix. $\operatorname{Diag}\{\mathbf{a}\}$ and $\operatorname{Diag}\{\mathbf{A}\}$ return diagonal matrices with the $i$-th main diagonal entry equal to the $i$-th entry of vector $\mathbf{a}$ and the $i$-th main diagonal entry of matrix $\mathbf{A}$, respectively. $\operatorname{diag}\{\mathbf{A}\}$ returns a vector with the $i$-th entry equal to the $i$-th main diagonal entry of matrix $\mathbf{A}$. $\operatorname{Re}\{\mathbf{A}\}$ returns a real-valued matrix whose entries equal the real parts of the entries of matrix $\mathbf{A}$. $\mathbf{I}_K$ denotes the $K$-by-$K$ identity matrix. $\left\|\mathbf{a}\right\|$ denotes the $l_2$-norm of vector $\mathbf{a}$. $\odot$ denotes the Hadamard product. $[\mathbf{A}]_{p_1,\cdots,p_K}$ denotes the entry with indices $p_1,\cdots,p_K$ of $K$-dimensional tensor $\mathbf{A}$. $\lfloor \cdot \rfloor$ denotes floor notation.
$\mathcal{CN}\left(\mathbf{0}, \mathbf{\Gamma}\right)$ denotes the circularly symmetric complex Gaussian (CSCG) distribution with mean $\mathbf{0}$ and covariance matrix $\mathbf{\Gamma}$.

\section{System Model and Problem Formulation}
\begin{figure}
	\centering
	\includegraphics[width=0.52\textwidth,height=0.45\textwidth]{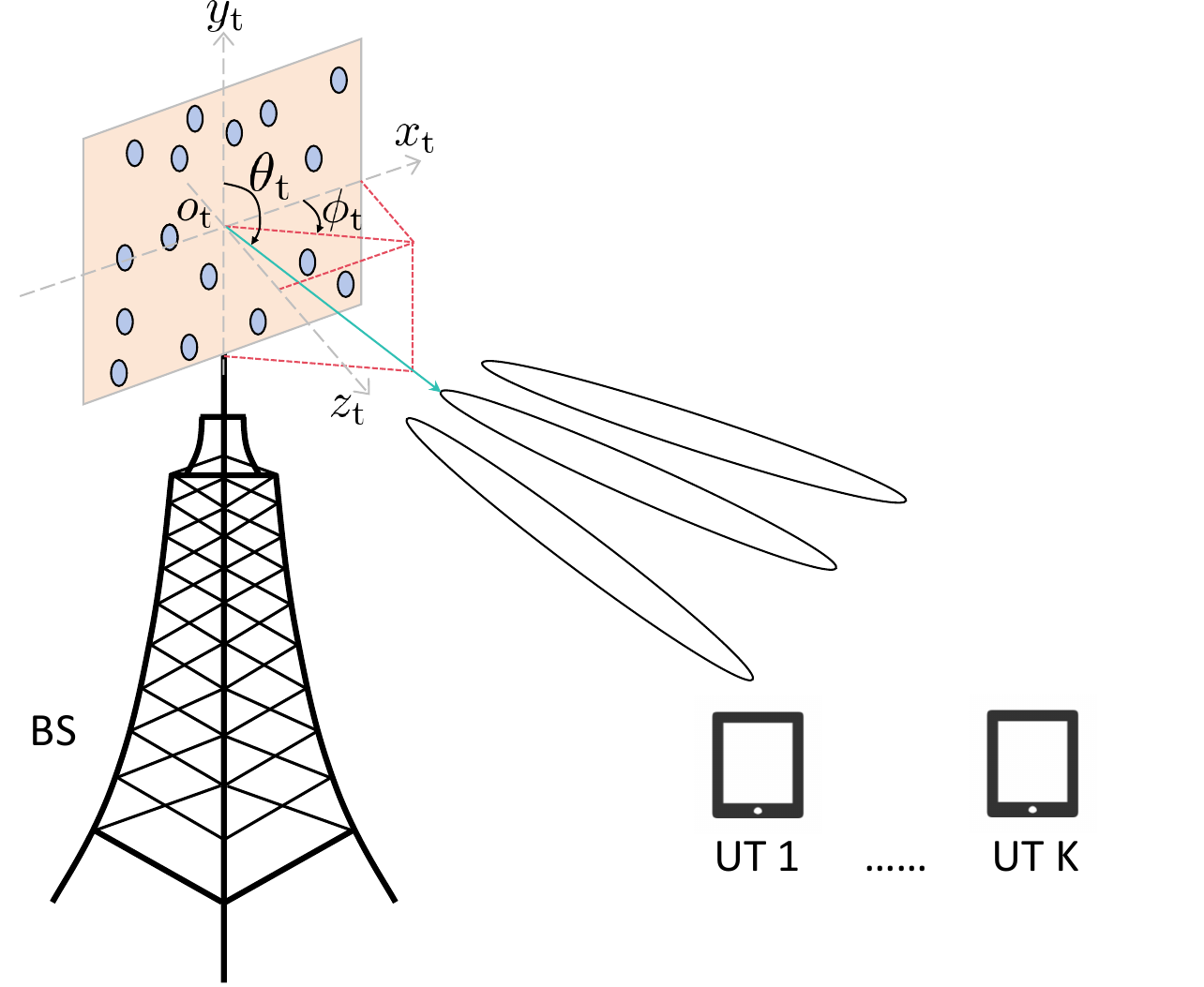}
	\caption{The MA-enhanced multi-user downlink communication system. The circles represent the transmit MAs. The graphical details of the UTs are similar to those of the BS and are omitted here for simplicity.}
	\label{fig: system diagram}
\end{figure}
As shown in Fig.~\ref{fig: system diagram}, we investigate a multi-user MIMO downlink communication system, where the BS and each of $K$ UTs are equipped with $N$ and $M$ MAs, respectively. The transmit MAs at the BS and the receive MAs at UT $k \in\{ 1,\cdots,K\}$ can move within square regions $\mathcal{C}_{\mathrm{t}}$ and $\mathcal{C}_{\mathrm{r},k}$, respectively. That is, the position of transmit MA $n$ and receive MA $m$ at the UT $k$ satisfy $\mathbf{t}_n \triangleq \left(x_{\mathrm{t},n},y_{\mathrm{t},n}\right) \in \mathcal{C}_{\mathrm{t}}$ and $\mathbf{r}_{k,m} \triangleq \left(x_{\mathrm{r},k,m},y_{\mathrm{r},k,m}\right) \in \mathcal{C}_{\mathrm{r},k}$, respectively. By stacking the coordinates of all MAs together, the transmit APV and the receive APV of UT $k$ are denoted by $\mathbf{t}\triangleq\left(\mathbf{x}_{\mathrm{t}};\mathbf{y}_{\mathrm{t}}\right)$ with $\mathbf{x}_{\mathrm{t}} \triangleq \left(x_{\mathrm{t},1},\cdots,x_{\mathrm{t},N}\right)^T$,
$\mathbf{y}_{\mathrm{t}} \triangleq \left(y_{\mathrm{t},1},\cdots,y_{\mathrm{t},N}\right)^T$ and $\mathbf{r}_k\triangleq\left(\mathbf{x}_{\mathrm{r},k};\mathbf{y}_{\mathrm{r},k}\right)$ with 
$\mathbf{x}_{\mathrm{r},k} \triangleq \left(x_{\mathrm{r},k,1},\cdots,x_{\mathrm{r},k,M}\right)^T$,
$\mathbf{y}_{\mathrm{r},k} \triangleq \left(y_{\mathrm{r},k,1},\cdots,y_{\mathrm{r},k,M}\right)^T$, respectively. To avoid potential coupling, a minimum distance $D \geq \lambda/2$ is required between each pair of MAs \cite{AAfL}, i.e., $\left\|\mathbf{t}_i-\mathbf{t}_j\right\| \geq D$, $\forall i \neq j$, and $\left\|\mathbf{r}_{k,i}-\mathbf{r}_{k,j}\right\| \geq D$, $\forall k$, $\forall i \neq j$, where $\lambda$ is the wavelength of the signal carrier. Then, the received signal at UT $k$ is given by
\begin{equation}
	\mathbf{y}_k=\mathbf{H}_k(\mathbf{t},\mathbf{r}_k) \sum_{i=1}^K \mathbf{P}_i \mathbf{s}_i+\mathbf{z}_k.
\end{equation}
Here, $\mathbf{H}_k\left(\mathbf{t},\mathbf{r}_k\right) \in \mathbb{C}^{M \times N}$ represents the channel matrix between the BS and UT $k$, which is related to the APVs $\mathbf{t}$ and $\mathbf{r}_k$. Assume that $s_k$ downlink streams are transmitted on each channel use. $\mathbf{P}_k \in \mathbb{C}^{N \times s_k}$ is a precoding matrix for UT $k$. The transmitted signal intended for UT $k$, $\mathbf{s}_k$, follows a zero-mean CSCG distribution with covariance matrix $\mathbb{E}\left\{\mathbf{s}_k \mathbf{s}_k^{H}\right\}=\mathbf{I}_{s_k}$ and satisfies $\mathbb{E}\left\{\mathbf{s}_k \mathbf{s}_{k'}^{H}\right\}=\mathbf{0}$, $\forall k' \neq k$. $\mathbf{z}_k \sim \mathcal{CN}\left(\mathbf{0}, \sigma^2 \mathbf{I}_{M}\right)$ denotes the CSCG noise with zero mean and variance $\sigma^2$ at UT $k$.

\subsection{Channel Model}
We consider the far-field channel model, and thus, the angles of departure (AoDs) and angles of arrival (AoAs) for different positions in transmit region $\mathcal{C}_{\mathrm{t}}$ and receive region $\mathcal{C}_{\mathrm{r},k}$ are identical, respectively \cite{Mapa,Mccf}. We assume the Rician channel model holds and there are $L_{\mathrm{t}}$ transmit paths and $L_{\mathrm{r}}$ receive paths overall. Let $(\theta_{\mathrm{t},k}^{l_{\mathrm{t}}}, \phi_{\mathrm{t},k}^{l_{\mathrm{t}}})$ and $(\theta_{\mathrm{r},k}^{l_{\mathrm{r}}}, \phi_{\mathrm{r},k}^{l_{\mathrm{r}}})$ denote the elevation and azimuth angles of the $l_{\mathrm{t}}$-th transmit path and the $l_{\mathrm{r}}$-th receive path between the BS and UT $k$, respectively. Then, the difference in the signal propagation distance between position $\mathbf{t}_n$ and the origin of $\mathcal{C}_{\mathrm{t}}$ at the BS and that between position $\mathbf{r}_{k,m}$ and the origin of $\mathcal{C}_{\mathrm{r},k}$ at UT $k$ are, respectively, given by \cite{Mapa,Mccf}
\begin{equation}
	\begin{aligned}
		\rho_{\mathrm{t},k}^{l_{\mathrm{t}}}(\mathbf{t}_n)&=x_{\mathrm{t},n} \sin \theta_{\mathrm{t},k}^{l_{\mathrm{t}}} \cos \phi_{\mathrm{t},k}^{l_{\mathrm{t}}}+y_{\mathrm{t},n} \cos \theta_{\mathrm{t},k}^{l_{\mathrm{t}}},\\
		\rho_{\mathrm{r},k}^{l_{\mathrm{r}}}(\mathbf{r}_{k,m})&=x_{\mathrm{r},k,m} \sin \theta_{\mathrm{r},k}^{l_{\mathrm{r}}} \cos \phi_{\mathrm{r},k}^{l_{\mathrm{r}}}+y_{\mathrm{r},k,m} \cos \theta_{\mathrm{r},k}^{l_{\mathrm{r}}}.
	\end{aligned}
\end{equation}
Then, the field response vectors of MA $n$ at the BS and MA $m$ at the UT $k$ are, respectively, given by
\begin{equation}
	\begin{aligned}
		\mathbf{g}_k(\mathbf{t}_n) \triangleq & \left[e^{j \frac{2 \pi}{\lambda} \rho_{\mathrm{t},k}^{1}(\mathbf{t}_n)}, \cdots, e^{j \frac{2 \pi}{\lambda} \rho_{\mathrm{t},k}^{L_{\mathrm{t}}}(\mathbf{t}_n)}\right]^{T},\\
		\mathbf{f}_{k}(\mathbf{r}_{k,m}) \triangleq & \left[e^{j \frac{2 \pi}{\lambda} \rho_{\mathrm{r},k}^{1}(\mathbf{r}_{k,m})},\cdots, e^{j \frac{2 \pi}{\lambda} \rho_{\mathrm{r},k}^{L_{\mathrm{r}}}(\mathbf{r}_{k,m})}\right]^{T}.
	\end{aligned} 
\end{equation}
Furthermore, the channel matrix $\mathbf{H}_k(\mathbf{t},\mathbf{r}_k)$ between the BS and UT $k$ can be written as
\begin{equation}
	\label{eq: discrete channel matrix}
	\begin{aligned}	
		\mathbf{H}_k(\mathbf{t},\mathbf{r}_k)
		= \mathbf{F}_{k}^H\left(\mathbf{r}_k\right) \mathbf{\Sigma}_k \mathbf{G}_{k}\left(\mathbf{t}\right),
	\end{aligned}
\end{equation}
where $\mathbf{F}_k(\mathbf{r}_k) \triangleq\left[\mathbf{f}_k\left(\mathbf{r}_{k,1}\right), \cdots, \mathbf{f}_k\left(\mathbf{r}_{k,M}\right)\right] \in \mathbb{C}^{L_{\mathrm{r}} \times M}$ and $\mathbf{G}_k(\mathbf{t}) \triangleq\left[\mathbf{g}_k\left(\mathbf{t}_1\right), \cdots, \mathbf{g}_k\left(\mathbf{t}_N\right)\right]\in \mathbb{C}^{L_{\mathrm{t}} \times N}$ denote the field response matrices at the BS and UT $k$, respectively. $\mathbf{\Sigma}_k\in\mathbb C^{L_{\mathrm{r}}\times L_{\mathrm{t}}}$ is the path response matrix, where $\left[\mathbf{\Sigma}_k\right]_{l_{\mathrm{r}},l_{\mathrm{t}}}$ represents the angular-domain gain of the $l_{\mathrm{t}}$-th transmit path and the $l_{\mathrm{r}}$-th receive path from the origin of transmit region $\mathcal{C}_{\mathrm{t}}$ to the origin of receive region $\mathcal{C}_{\mathrm{r},k}$ \cite{MCwm,WSRM}.

In MA-enhanced systems, exploiting I-CSI to design the APVs is usually impractically costly due to the constant and dynamic changes in the physical environment. Consequently, this paper employs the relatively stable S-CSI for designing the APVs. Without loss of generality, we assume that the LOS path is in the $1$-st AOD and $1$-st AOA. Thus, the path response matrix can be given by
\begin{equation}
	\begin{aligned}
		\mathbf{\Sigma}_k = \overline{\mathbf{\Sigma}}_k + \widetilde{\mathbf{\Sigma}}_k,
	\end{aligned}
\end{equation}
where $\overline{\mathbf{\Sigma}}_k$ and $\widetilde{\mathbf{\Sigma}}_k$ are the path response matrices associated with the LOS path and the Non-LOS (NLOS) paths, respectively.
Note that $\overline{\mathbf{\Sigma}}_k=\mathbf{0}$ if Rayleigh channels are considered; Otherwise, $\left[\overline{\mathbf{\Sigma}}_k\right]_{1,1}\neq0,\left[\overline{\mathbf{\Sigma}}_k\right]_{i,j}=0,i\neq1 \text{ or } j\neq1$ if Rician channels are considered. Besides, matrix $\widetilde{\mathbf{\Sigma}}_k$ can be written as
\begin{equation}
	\begin{aligned}
		\widetilde{\mathbf{\Sigma}}_k = \mathbf{M}_k \odot  \mathbf{W}_k,
	\end{aligned}
\end{equation}
where the entries of $\mathbf{M}_k$ are non-negative and represent the average path gain and the entries of $\mathbf{W}_k$ are independently and identically distributed (i.i.d.) zero-mean complex Gaussian random variables with unit variance. 

\subsection{Problem Formulation}
We aim to maximize the system EE, which is formulated as the ratio of the achievable sum rate to the corresponding power consumption. Thus, we first introduce the achievable rate model and the power consumption model, followed by the formulated optimization problem.

Given that the BS has access to only S-CSI, we focus on the average achievable rate. Furthermore, the optimal combining is employed after UT $k$ acquires knowledge of the I-CSI of its own channel through appropriately designed pilot signals \cite{Bdma}, along with the aggregate instantaneous interference-plus-noise covariance matrix. Under these premises, the average achievable rate for UT $k$ can be expressed as \cite{Rtfm}, \cite{Bdst} 
\begin{equation}
	\label{eq: Rk}
	\begin{aligned}
		R_k\left(\mathbf{t},\mathbf{r}_k,\mathbf{P}\right) &= \mathbb{E}_{\widetilde{\mathbf{\Sigma}}_k}\left\{\log 	\operatorname{det}\left(\mathbf{I}_M+\frac{1}{\sigma^2} \mathbf{H}_k \mathbf{P} \mathbf{P}^H \mathbf{H}_k^H\right)\right. \\
		& \left.-\log 	\operatorname{det}\left(\mathbf{I}_M+\frac{1}{\sigma^2} \mathbf{H}_k \mathbf{P}_{\backslash k} \mathbf{P}_{\backslash k}^H \mathbf{H}_k^H\right)\right\},
	\end{aligned}
\end{equation}
where $\mathbf{P}=\left[\mathbf{P}_1, \mathbf{P}_2, \cdots, \mathbf{P}_K\right]$ denotes the matrix obtained by stacking precoding matrices of all UTs horizontally, $\mathbf{P}_{\backslash k}=\left[\mathbf{P}_1, \mathbf{P}_2, \cdots, \mathbf{P}_{k-1}, \mathbf{P}_{k+1}, \cdots, \mathbf{P}_K\right]$, and $\mathbb{E}_{\widetilde{\mathbf{\Sigma}}_k}$ denotes the expectation takes over I-CSI. In addition, we consider an affine power consumption model \cite{Eeof}. In particular, the total power consumed is denoted as
\begin{equation}
	P_{\mathrm{tot}}=\omega \operatorname{tr}\left(\mathbf{P} \mathbf{P}^H\right)+N P_{\mathrm{c}}+P_{\mathrm{s}},
\end{equation}
where $\omega$ describes the transmit power amplifier efficiency, $\operatorname{tr}\left(\mathbf{P} \mathbf{P}^H\right)$ is the total transmit power, $P_{\mathrm{c}}$ denotes the dynamic power consumption associated with each antenna (e.g., circuit power of corresponding radio frequency (RF) signal processing and antenna movement, which is always proportional to $N$ and independent of $\operatorname{tr}\left(\mathbf{P} \mathbf{P}^H\right)$), and $P_{\mathrm{s}}$ accounts for the static circuit power consumption, which is independent of both $N$ and $\operatorname{tr}\left(\mathbf{P} \mathbf{P}^H\right)$.

In this paper, we aim to jointly optimize the transmit/receive APVs and the precoding matrix under the position constraints and the power constraint. Consequently, the optimization problem is formulated as follows
\begin{equation}
	\label{op: original}
	\begin{aligned}
		\max_{\mathbf{t}, \left\{\mathbf{r}_k\right\}, \mathbf{P}} \quad & \frac{ \sum_{k=1}^K R_k\left(\mathbf{t},\mathbf{r}_k,\mathbf{P}\right) } {P_\mathrm{tot}} \\
		\text { s.t. } \quad
		&\text{C1: }\mathbf{t}_n\in \mathcal{C}_{\mathrm{t}}, \forall n,\\
		&\text{C2: }\left\|\mathbf{t}_i-\mathbf{t}_j\right\| \geq D,  \forall i \neq j,\\
		&\text{C3$_k$: }\mathbf{r}_{k,m} \in \mathcal{C}_{\mathrm{r},k}, \forall m, &&\forall k, \\
		&\text{C4$_k$: }\left\|\mathbf{r}_{k,i}-\mathbf{r}_{k,j}\right\| \geq D, \forall i \neq j, \,\,\,\, &&\forall k, \\
		&\text{C5: } \operatorname{tr}\left(\mathbf{P}\mathbf{P}^H\right) \leq P_\mathrm{max},
	\end{aligned}
\end{equation}
where $P_\mathrm{max} \geq 0$ is the given maximum available transmit power.
Note that problem \eqref{op: original} is challenging to obtain the optimal solution for the following reasons. Firstly, the achievable rates in the objective function involve expectations and are highly non-convex in terms of the APVs and precoding matrices. Secondly, the position constraints of the MAs, C2 and C4$_k$, $\forall k$, are non-convex. As a result, we focus on acquiring a low-complexity solution by alternatively optimizing the transmit variables and receive APVs through an AO algorithm combing the DE and SCA approaches.

\section{Multi-user MA-Enhanced MIMO System}
\label{section: Multi-user MA-Enhanced MIMO System}
\begin{figure*}[!h]
	\centering
	\includegraphics[width=0.85\textwidth,height=0.36\textwidth]{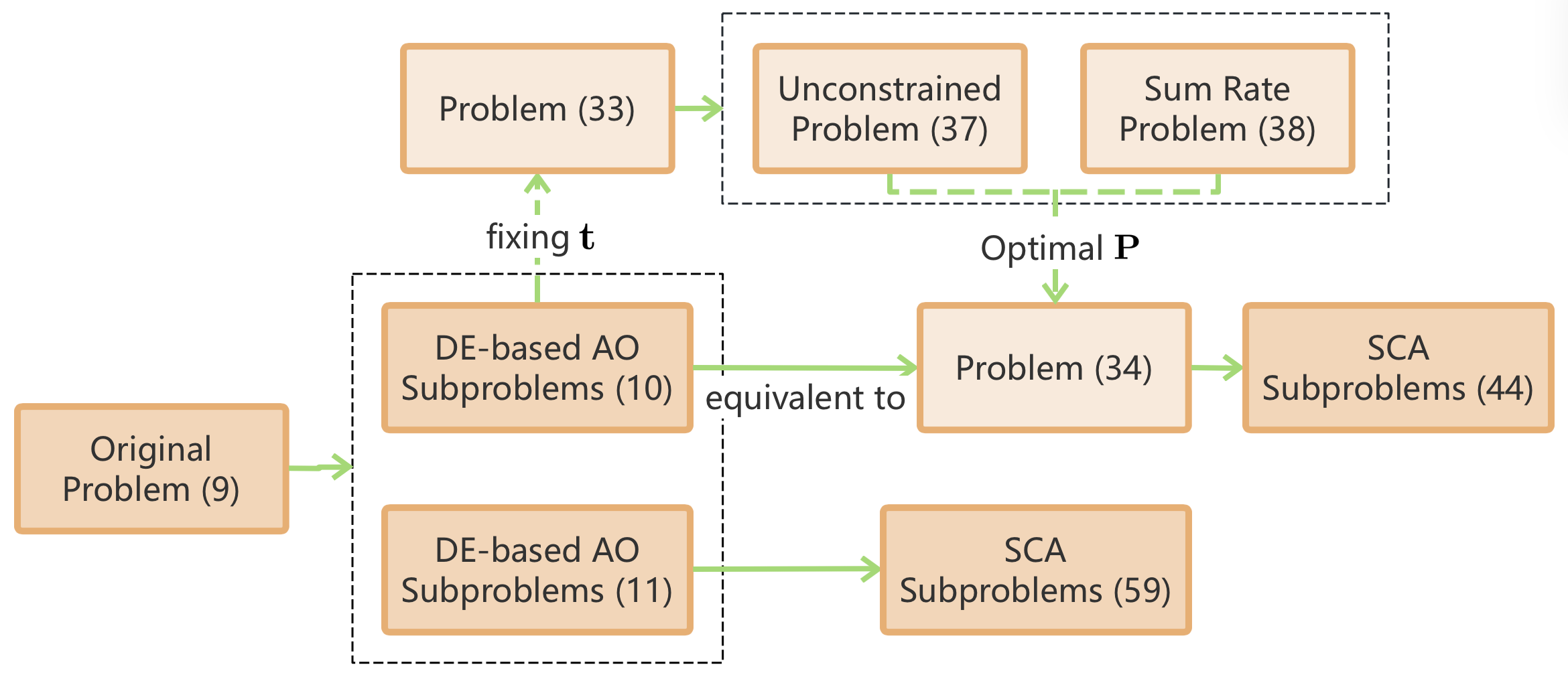}
	\caption{Illustration of the AO algorithm framework.}
	\label{fig: algorithm_framework}
\end{figure*}
In this section, we first sketch the AO algorithm framework, which alternatively optimizes the transmit variables and the receive APVs to maximize the deterministic objective functions formulated by the DE technique. Then, we introduce the formulation of the deterministic objective functions, i.e., the deterministic minorizing function of the average achievable sum rate and the deterministic function of each UT's average achievable rate w.r.t. the transmit variables and the corresponding receive APV, respectively. Finally, we present the low-complexity SCA algorithms for solving the subproblems in each iteration of the AO algorithm. The overall AO algorithm framework for EE maximization problem \eqref{op: original} is summarized in Fig.~\ref{fig: algorithm_framework}, which will be described in details in the following.

\subsection{AO Algorithm Framework}
In iteration $\ell$ of our proposed AO algorithm, we first construct deterministic minorizing functions $\left\{\hat{R}_{\mathrm{t},k}^{\ell}\left(\mathbf{t},\mathbf{P}\right)\right\}$ by the DE technique.
Then, we optimize $\mathbf{t}$ and $\mathbf{P}$, to maximize the deterministic minorizing function, $EE^{\ell}\left(\mathbf{t},\mathbf{P}\right)\triangleq\frac{ \sum_{k=1}^K \hat{R}_{\mathrm{t},k}^{\ell}\left(\mathbf{t},\mathbf{P}\right) } {P_\mathrm{tot}}$, via surrogate problem given by
\begin{equation}
	\label{op: EE t P}
	\begin{aligned}
		\max_{\mathbf{t},\mathbf{P}} \quad & EE^{\ell}\left(\mathbf{t},\mathbf{P}\right) \\
		\text { s.t. } \quad &\text{C1},\text{C2},\text{C5}.
	\end{aligned}
\end{equation}
Subsequently, we construct the deterministic functions of the average achievable rates, $\left\{\bar{R}_{\mathrm{r},k}\left(\mathbf{r}_k\right)\right\}$.
Finally, a series of deterministic functions, $\left\{\bar{R}_{\mathrm{r},k}\left(\mathbf{r}_k\right)\right\}$, are maximized by optimizing the receive APVs, $\left\{\mathbf{r}_k\right\}$, respectively, as follows
\begin{equation}
	\label{op: SR r}
	\begin{aligned}
		\max_{\mathbf{r}_k} \quad & \bar{R}_{\mathrm{r},k}\left(\mathbf{r}_k\right) \\
		\text { s.t. } \quad &\text{C3}_k,\text{C4}_k.
	\end{aligned}
\end{equation}
Next, we will present the formulation of functions $\hat{R}_{\mathrm{t},k}^{\ell}\left(\mathbf{t},\mathbf{P}\right)$, $\bar{R}_{\mathrm{r},k}\left(\mathbf{r}_k\right)$, $\forall k$.

\subsection{Deterministic Function Formulation}
\label{Section: Formulating deterministic functions}
For simplicity of notations, we consider the formulation in iteration $\ell+1$ of the AO algorithm. To formulate $\hat{R}_{\mathrm{t},k}^{\ell+1}\left(\mathbf{t},\mathbf{P}\right)$, $\forall k$, we first give a minorizing function w.r.t. the transmit variables of average achievable rate $R_k\left(\mathbf{t},\mathbf{r}_k,\mathbf{P}\right)$ and then derive the DEs of the parameters needed for the minorizing function.

Given $\mathbf{t}^{\ell}$, $\left\{\mathbf{r}_k^{\ell}\right\}$ and $\left\{\mathbf{P}_k^{\ell}\right\}$, let the covariance matrix of the total received and interference signals at the $k$-th UT be defined as
\begin{equation}
	\begin{aligned}
		\check{\mathbf{K}}_k^{\ell}&=\sigma^2 \mathbf{I}_{M}+\mathbf{H}_k^{\ell} \mathbf{P}^{\ell}(\mathbf{P}^{\ell})^H \left(\mathbf{H}_k^{\ell}\right)^H,\\
		\mathbf{K}_k^{\ell}&=\sigma^2 \mathbf{I}_{M}+\mathbf{H}_k^{\ell} \mathbf{P}_{\backslash k}^{\ell}(\mathbf{P}_{\backslash k}^{\ell})^H \left(\mathbf{H}_k^{\ell}\right)^H,
	\end{aligned}
\end{equation}
where $\mathbf{H}_k^{\ell}=\mathbf{F}_k\left( \mathbf{r}_k^{\ell} \right)^H \mathbf{\Sigma}_k \mathbf{G}_k \left( \mathbf{t}^{\ell} \right)$, $\mathbf{P}^{\ell}=\left[\mathbf{P}_1^{\ell}, \mathbf{P}_2^{\ell}, \cdots, \mathbf{P}_K^{\ell}\right]$ and $\mathbf{P}_{\backslash k}^{\ell}=\left[\mathbf{P}_1^{\ell}, \mathbf{P}_2^{\ell}, \cdots, \mathbf{P}_{k-1}^{\ell}, \mathbf{P}_{k+1}^{\ell}, \cdots, \mathbf{P}_K^{\ell}\right]$.
Then, the following lemma gives a minorizing function of the average achievable rate \cite{SROS}:
\begin{lemma}
	\label{lemma: minorizing function}
	Let $R_{\mathrm{t},k}^{\ell+1}$ be a function given as
	\begin{equation}
		\label{eq: minorizing function t}
		\begin{aligned}
			R_{\mathrm{t},k}^{\ell+1}\left(\mathbf{t},\mathbf{P}\right) = &	\operatorname{tr}\left(\left(\mathbf{A}_k^{\ell}\right)^H \mathbf{G}_k \left( \mathbf{t} \right)\mathbf{P}_k\right) + \operatorname{tr}\left(\mathbf{A}_k^{\ell} \mathbf{P}_k^H\mathbf{G}_k\left( \mathbf{t} \right)^H \right) \\-&\operatorname{tr}\left(\mathbf{B}_k^{\ell} \mathbf{G}_k \left( \mathbf{t} \right) \mathbf{P} \mathbf{P}^H \mathbf{G}_k\left( \mathbf{t} \right)^H \right) + c_k^{\ell},
		\end{aligned}
	\end{equation}
	where matrices $\mathbf{A}_k^{\ell}$, $\mathbf{B}_k^{\ell}$ and constants $c_k^{\ell}$ are given in \eqref{eq: minorizing function ABctk} at the bottom of this page.
	\begin{figure*}[!b]
		\hrule
		\begin{subequations}
			\label{eq: minorizing function ABctk}
			\begin{align}
				\label{eq: minorizing function Atk}
				\mathbf{A}_k^{\ell} & 		=\mathbb{E}_{\widetilde{\mathbf{\Sigma}}_k}\left\{  \mathbf{\Sigma}_k^H \mathbf{F}_k \left( \mathbf{r}_k^{\ell} \right) \left(\mathbf{K}_k^{\ell}\right)^{-1} \mathbf{F}_k\left( \mathbf{r}_k^{\ell} \right)^H \mathbf{\Sigma}_k \right\} \mathbf{G}_k \left( \mathbf{t}^{\ell} \right)\mathbf{P}_k^{\ell}, \\
				\label{eq: minorizing function Btk}
				\mathbf{B}_k^{\ell} & 		=\mathbb{E}_{\widetilde{\mathbf{\Sigma}}_k}\left\{  \mathbf{\Sigma}_k^H \mathbf{F}_k \left( \mathbf{r}_k^{\ell} \right) \left(\left(\mathbf{K}_k^{\ell}\right)^{-1}-\left(\check{\mathbf{K}}_k^{\ell}\right)^{-1}\right) \mathbf{F}_k\left( \mathbf{r}_k^{\ell} \right)^H \mathbf{\Sigma}_k \right\},\\
				\label{eq: minorizing function ctk}
				c_k^{\ell} &
				= s_k + \mathbb{E}_{\widetilde{\mathbf{\Sigma}}_k}\left\{\log 	 \operatorname{det}\left(\mathbf{C}_k^{\ell}\right) - \operatorname{tr}\left(\mathbf{C}_k^{\ell}\right)
				-\sigma^2 	\operatorname{tr}\left(\mathbf{C}_k^{\ell}\left(\mathbf{D}_k^{\ell}\right)^H  \mathbf{D}_k^{\ell}\right)\right\}, \\
				\mathbf{C}_k^{\ell}&=\mathbf{I}_{s_k}+\left(\mathbf{P}_k^{\ell}\right)^H \left(\mathbf{H}_k^{\ell}\right)^H \left(\mathbf{K}_k^{\ell}\right)^{-1} \mathbf{H}_k^{\ell} \mathbf{P}_k^{\ell}, \mathbf{D}_k^{\ell}=\left(\check{\mathbf{K}}_k^{\ell}\right)^{-1}\mathbf{H}_k^{\ell}\mathbf{P}_k^{\ell}.
			\end{align}
		\end{subequations}
	\end{figure*}
	Then it is a minorizing function w.r.t. the transmit variables of average achievable rate $R_k\left(\mathbf{t},\mathbf{r}_k,\mathbf{P}\right)$,  which satisfies the following conditions:
	\begin{subequations}
		\begin{align}
			R_{\mathrm{t},k}^{\ell+1}\left(\mathbf{t},\mathbf{P}\right) &\leq R_k\left(\mathbf{t},\mathbf{r}_k,\mathbf{P}\right), \\
			\label{eq: value equality}
			R_{\mathrm{t},k}^{\ell+1}\left(\mathbf{t}^{\ell},\mathbf{P}^{\ell}\right) &= R_k\left(\mathbf{t}^{\ell},\mathbf{r}_k^{\ell},\mathbf{P}^{\ell}\right), \\
			\nabla_{\mathbf{t}} R_{\mathrm{t},k}^{\ell+1}\left(\mathbf{t}^{\ell},\mathbf{P}^{\ell}\right) &= \nabla_{\mathbf{t}} R_k\left(\mathbf{t}^{\ell},\mathbf{r}_k^{\ell},\mathbf{P}^{\ell}\right), \\
			\nabla_{\mathbf{P}} R_{\mathrm{t},k}^{\ell+1}\left(\mathbf{t}^{\ell},\mathbf{P}^{\ell}\right) &= \nabla_{\mathbf{P}} R_k\left(\mathbf{t}^{\ell},\mathbf{r}_k^{\ell},\mathbf{P}^{\ell}\right).
		\end{align}
	\end{subequations}
\end{lemma}
  
While the expectations needed in constructing minorizing function \eqref{eq: minorizing function t} can be acquired via the MC method, it is still of high computational complexity. Therefore, we exploit the DE approach to equivalently establish a deterministic minorizing function \cite{SROS}. The DE technique is a methodology developed from operator-valued free probability theory, which can be leveraged to provide an efficient analytical framework for evaluating the average achievable rates. This technique constructs closed-form approximations by solving iteratively convergent equations, thereby circumventing direct MC averaging over massive high-dimensional random states. We begin with the deterministic function of the positive term in average achievable rate $R_k\left(\mathbf{t},\mathbf{r}_k,\mathbf{P}\right)$ in \eqref{eq: Rk} w.r.t. the transmit variables, which is given as \cite{FDEf}
\begin{equation}
	\label{eq: DE Rtk+}
	\begin{aligned}
		\bar{R}_{\mathrm{t},k}^+\left(\mathbf{t},\mathbf{P}\right) =& \log \operatorname{det}\left( \mathbf{I}_{L_{\mathrm{t}}}+ \widetilde{\mathbf{\Gamma}}_k^+ \mathbf{G}_k \left( \mathbf{t} \right) \mathbf{P} \mathbf{P}^H \mathbf{G}_k\left( \mathbf{t} \right)^H  \right) \\
		+& \log \operatorname{det}\left(\widetilde{\mathbf{\Phi}}_k^+\right) - \operatorname{tr}\left( \sigma^2 \left( \mathbf{I}_M - \widetilde{\mathbf{\Phi}}_k^+ \right) \widetilde{\mathbf{\Theta}}_k^+ \right),
	\end{aligned}
\end{equation}
where matrix $\widetilde{\mathbf{\Gamma}}_k^+ = \sigma^{-2} \overline{\mathbf{\Sigma}}_k^H \mathbf{F}_k\left( \mathbf{r}_k \right) \left(\widetilde{\mathbf{\Phi}}_k^+\right)^{-1} \mathbf{F}_k\left( \mathbf{r}_k \right)^H \overline{\mathbf{\Sigma}}_k - \eta_k^+ \left( \widetilde{\mathbf{\Theta}}_k^+ \right)$.
The other matrices are calculated through the following iterative process:
\begin{subequations}
	\label{eq: DE process}
	\begin{align}
		\label{eq: DE proces (a)}
		& \widetilde{\mathbf{\Phi}}_k^+ = \mathbf{I}_M - \mathbf{F}_k\left( \mathbf{r}_k \right)^H \tilde{\eta}_k^+ \left( \mathbf{\Theta}_k^+ \right) \mathbf{F}_k\left( \mathbf{r}_k \right),
		\\
		\label{eq: DE proces (b)}
		& \mathbf{\Phi}_k^+ = \mathbf{I}_N - \mathbf{Q}^{\frac{H}{2}} \mathbf{G}_k\left( 	\mathbf{t} \right)^H  \eta_k^+ \left( \widetilde{\mathbf{\Theta}}_k^+ \right) \mathbf{G}_k \left( \mathbf{t} \right) \mathbf{Q}^{\frac{1}{2}}, \\
		\label{eq: DE proces (c)}
		& \widetilde{\mathbf{\Theta}}_k^+ = \left( -\sigma^2 \widetilde{\mathbf{\Phi}}_k^+ - \overline{\mathbf{H}}_k \mathbf{Q}^{\frac{1}{2}} \left(\mathbf{\Phi}_k^+\right)^{-1} \mathbf{Q}^{\frac{H}{2}} \overline{\mathbf{H}}_k^H \right)^{-1}, \\
		\label{eq: DE proces (d)}
		\nonumber
		& \mathbf{\Theta}_k^+ = \left( -\sigma^2 \mathbf{\Phi}_k^+ -  	\mathbf{Q}^{\frac{H}{2}} \overline{\mathbf{H}}_k^H \left(\widetilde{\mathbf{\Phi}}_k^+\right)^{-1} \overline{\mathbf{H}}_k \mathbf{Q}^{\frac{1}{2}} \right)^{-1} \\
		&\overset{(a)}{=} -\sigma^{-2}\left( \mathbf{\Phi}_k^+ \right)^{-1} \left( \mathbf{I}_N + \mathbf{Q}^{\frac{H}{2}} \overline{\mathbf{H}}_k^H \widetilde{\mathbf{\Theta}}_k^+ \overline{\mathbf{H}}_k \mathbf{Q}^{\frac{1}{2}} \left( \mathbf{\Phi}_k^+ \right)^{-1} \right),
	\end{align}
\end{subequations}
where $\mathbf{Q}=\mathbf{P}\mathbf{P}^H$ is the covariance matrix of the transmitted signal intended for all UTs; matrix-valued functions $\tilde{\eta}_k^+$ and $\eta_k^+$ are written as \eqref{eq: matrix-valued function 1} and \eqref{eq: matrix-valued function 2}, respectively; when $L_{\mathrm{t}}\leq N$, the inverse of $\mathbf{\Phi}_k^+$ can be calculated by \eqref{eq: Phi inv} according to the Woodbury identity to reduce coomputational complexity; ($a$) holds also due to the Woodbury identity \cite{TMCb}.
\begin{figure*}
	\begin{subequations}
		\begin{align}
			\label{eq: matrix-valued function 1}
			\tilde{\eta}_k^+ \left( \mathbf{\Theta}_k^+ \right) &= \mathbb{E}_{\widetilde{\mathbf{\Sigma}}_k} \left\{ \widetilde{\mathbf{\Sigma}}_k \mathbf{G}_k \left( \mathbf{t} \right) \mathbf{Q}^{\frac{1}{2}} \mathbf{\Theta}_k^+ \mathbf{Q}^{\frac{H}{2}} \mathbf{G}_k\left( \mathbf{t} \right)^H  \widetilde{\mathbf{\Sigma}}_k^H \right\}
			= \operatorname{Diag} \left\{ \mathbf{M}_k \operatorname{Diag} \left\{	\mathbf{G}_k \left( \mathbf{t} \right) \mathbf{Q}^{\frac{1}{2}} \mathbf{\Theta}_k^+ \mathbf{Q}^{\frac{H}{2}} \mathbf{G}_k\left( \mathbf{t} \right)^H  \right\} \mathbf{M}_k^H \right\}, \\
			\label{eq: matrix-valued function 2}
			\eta_k^+ \left( \widetilde{\mathbf{\Theta}}_k^+ \right) &= \mathbb{E}_{\widetilde{\mathbf{\Sigma}}_k} \left\{ \widetilde{\mathbf{\Sigma}}_k^H \mathbf{F}_k\left( \mathbf{r}_k \right) \widetilde{\mathbf{\Theta}}_k^+ \mathbf{F}_k\left( \mathbf{r}_k \right)^H \widetilde{\mathbf{\Sigma}}_k \right\}
			= \operatorname{Diag} \left\{ \mathbf{M}_k^H \operatorname{Diag} \left\{ \mathbf{F}_k\left( \mathbf{r}_k \right) \widetilde{\mathbf{\Theta}}_k^+ \mathbf{F}_k\left( \mathbf{r}_k \right)^H \right\} \mathbf{M}_k \right\}. \\
			\label{eq: Phi inv}
			\left( \mathbf{\Phi}_k^+ \right)^{-1}&= \mathbf{I}_N + \mathbf{Q}^{\frac{H}{2}} \mathbf{G}_k\left( \mathbf{t} \right)^H \left(\left( \eta_k^+ \left( \widetilde{\mathbf{\Theta}}_k^+ \right)\right)^{-1} - \mathbf{G}_k \left( \mathbf{t} \right) \mathbf{Q} \mathbf{G}_k\left( 	\mathbf{t} \right)^H \right)^{-1} \mathbf{G}_k \left( \mathbf{t} \right) \mathbf{Q}^{\frac{1}{2}}.
		\end{align}
	\end{subequations}
	\hrule
\end{figure*}
Moreover, the deterministic function of the negative term in average achievable rate $R_k\left(\mathbf{t},\mathbf{r}_k,\mathbf{P}\right)$ in \eqref{eq: Rk} w.r.t. the transmit variables is formulated as
\begin{equation}
	\label{eq: DE Rtk-}
	\begin{aligned}
		\bar{R}_{\mathrm{t},k}^-(\mathbf{t},\mathbf{P}_{\backslash k}) =& \log \operatorname{det}\left( \mathbf{I}_{L_{\mathrm{t}}} + \widetilde{\mathbf{\Gamma}}_k^- \mathbf{G}_k \left( \mathbf{t} \right) \mathbf{P}_{\backslash k} \mathbf{P}_{\backslash k}^H \mathbf{G}_k\left( \mathbf{t} \right)^H  \right) \\
		+& \log \operatorname{det}\left(\widetilde{\mathbf{\Phi}}_k^-\right) - \operatorname{tr}\left( \sigma^2 \left( \mathbf{I}_M - \widetilde{\mathbf{\Phi}}_k^- \right) \widetilde{\mathbf{\Theta}}_k^- \right),
	\end{aligned}
\end{equation}
with matrix $\widetilde{\mathbf{\Gamma}}_k^- = \sigma^{-2} \overline{\mathbf{\Sigma}}_k^H \mathbf{F}_k\left( \mathbf{r}_k \right) \left(\widetilde{\mathbf{\Phi}}_k^-\right)^{-1} \mathbf{F}_k\left( \mathbf{r}_k \right)^H \overline{\mathbf{\Sigma}}_k - \eta_k^- \left( \widetilde{\mathbf{\Theta}}_k^- \right)$. Here, an iterative procedure similar to \eqref{eq: DE process} with $\mathbf{Q}$ being replaced by $\mathbf{Q}_{\backslash k}=\mathbf{P}_{\backslash k}\mathbf{P}_{\backslash k}^H$ is adopted to generate matrices $\widetilde{\mathbf{\Phi}}_k^-$, $\mathbf{\Phi}_k^-$, $\mathbf{\Theta}_k^-$ and $\widetilde{\mathbf{\Theta}}_k^-$.

Given variables $\mathbf{t}^{\ell}$, $\left\{\mathbf{r}_k^{\ell}\right\}$ and $\left\{\mathbf{P}_k^{\ell}\right\}$, the deterministic functions of the average achievable rates, $\bar{R}_{\mathrm{t},k}\left(\mathbf{t},\mathbf{P}\right)\triangleq \bar{R}_{\mathrm{t},k}^+\left(\mathbf{t},\mathbf{P}\right)-\bar{R}_{\mathrm{t},k}^-(\mathbf{t},\mathbf{P}_{\backslash k})$, can be constructed easily based on the S-CSI. To calculate the DEs of constants $\left\{c_k^{\ell}\right\}$ and matrices $\left\{\mathbf{A}_k^{\ell}\right\}$, $\left\{\mathbf{B}_k^{\ell}\right\}$, we need to compare the derivatives of $R_k(\mathbf{t},\mathbf{r}_k,\mathbf{P})$ and $\bar{R}_{\mathrm{t},k}\left(\mathbf{t},\mathbf{P}\right)$ w.r.t. $\left(\mathbf{G}_k \left( \mathbf{t}\right)\mathbf{P}_k\right)^H$ and $\left(\mathbf{G}_l \left( \mathbf{t}\right)\mathbf{P}_k\right)^H$, $l\neq k$.
We begin with the derivatives of $R_k(\mathbf{t},\mathbf{r}_k,\mathbf{P})$.
From \eqref{eq: Rk} and 
\eqref{eq: minorizing function Btk}, differentiating $R_k(\mathbf{t},\mathbf{r}_k,\mathbf{P})$ gives us
\begin{equation}
	\begin{aligned}
		\label{eq: diff Rk GP}
		&\left.\frac{\partial R_k(\mathbf{t},\mathbf{r}_k,\mathbf{P})}{\partial \left(\mathbf{G}_k \left( \mathbf{t}\right)\mathbf{P}_k\right)^H}\right|_{\mathbf{t}=\mathbf{t}^{\ell},\mathbf{r}_k=\mathbf{r}_k^{\ell},\mathbf{P}_k=\mathbf{P}_k^{\ell}}=\\
		&\mathbb{E}_{\widetilde{\mathbf{\Sigma}}_k}\left\{\mathbf{\Sigma}_k^H\mathbf{F}_k \left( \mathbf{r}_k^{\ell} \right) \left(\check{\mathbf{K}}_k^{\ell}\right)^{-1} \mathbf{F}_k\left( \mathbf{r}_k^{\ell} \right)^H \mathbf{\Sigma}_k \right\} \mathbf{G}_k \left( \mathbf{t}^{\ell} \right) \mathbf{P}_k^{\ell}
	\end{aligned}
\end{equation}
as well as
\begin{equation}
	\begin{aligned}
		\label{eq: Btl}
		&\left.\frac{\partial 	R_l(\mathbf{t},\mathbf{r}_l,\mathbf{P})}{\partial \left(\mathbf{G}_l \left( \mathbf{t} \right)\mathbf{P}_k\right)^H}\right|_{\mathbf{t}=\mathbf{t}^{\ell},\mathbf{r}_k=\mathbf{r}_k^{\ell},\mathbf{P}_k=\mathbf{P}_k^{\ell}}=-\mathbf{B}_l^{\ell} \mathbf{G}_l \left( \mathbf{t}^{\ell} \right) \mathbf{P}_k^{\ell}, \forall l\neq k.
	\end{aligned}
\end{equation}
Observing \eqref{eq: minorizing function Atk}, \eqref{eq: minorizing function Btk} and \eqref{eq: diff Rk GP}, the relation between $\mathbf{A}_k^{\ell}$ and $\mathbf{B}_k^{\ell}$ is given as
\begin{equation}
	\label{eq: Atk}
	\begin{aligned}
		\mathbf{A}_k^{\ell}	=&\left.\frac{\partial R_k(\mathbf{t},\mathbf{r}_k,\mathbf{P})}{\partial \left(\mathbf{G}_k \left( \mathbf{t}\right)\mathbf{P}_k\right)^H}\right|_{\mathbf{t}=\mathbf{t}^{\ell},\mathbf{r}_k=\mathbf{r}_k^{\ell},\mathbf{P}_k=\mathbf{P}_k^{\ell}}+\mathbf{B}_k^{\ell}\mathbf{G}_k \left( \mathbf{t}^{\ell} \right) \mathbf{P}_k^{\ell}.
	\end{aligned}
\end{equation}
Furthermore, according to \eqref{eq: value equality}, constant $c_k^{\ell}$ can be calculated as 
\begin{equation}
	\label{eq: ctk}
	\begin{aligned}
		c_k^{\ell} &= R_k\left(\mathbf{t}^{\ell},\mathbf{r}_k^{\ell},\mathbf{P}^{\ell}\right) +\operatorname{tr}\left(\mathbf{B}_k^{\ell} \mathbf{G}_k \left( \mathbf{t}^{\ell} \right) \mathbf{P}^{\ell} \left(\mathbf{P}^{\ell}\right)^H \mathbf{G}_k\left( \mathbf{t}^{\ell} \right)^H \right) \\&- \operatorname{tr}\left(\left(\mathbf{A}_k^{\ell}\right)^H \mathbf{G}_k \left( \mathbf{t}^{\ell} \right)\mathbf{P}_k^{\ell}\right) - \operatorname{tr}\left(\mathbf{A}_k^{\ell} \left(\mathbf{P}_k^{\ell}\right)^H\mathbf{G}_k\left( \mathbf{t}^{\ell} \right)^H \right).
	\end{aligned}
\end{equation}

Now, by observing \eqref{eq: DE Rtk+} and \eqref{eq: DE Rtk-}, we find that $\bar{R}_{\mathrm{t},k}\left(\mathbf{t},\mathbf{P}\right)$ are only related to the precoding matrices and field response matrices. Differentiating $\bar{R}_{\mathrm{t},k}\left(\mathbf{t},\mathbf{P}\right)$ w.r.t. $\left(\mathbf{G}_k \left( \mathbf{t}\right)\mathbf{P}_k\right)^H$ and $\left(\mathbf{G}_l \left( \mathbf{t}\right)\mathbf{P}_k\right)^H$, we obtain
\begin{equation}
	\begin{aligned}
		\label{eq: diff DE Rk GP}
		& \left.\frac{\partial 	\bar{R}_{\mathrm{t},k}\left(\mathbf{t},\mathbf{P}\right)}{\partial \left(\mathbf{G}_k \left( \mathbf{t}\right)\mathbf{P}_k\right)^H}\right|_{\mathbf{t}=\mathbf{t}^{\ell},\mathbf{r}_k=\mathbf{r}_k^{\ell},\mathbf{P}_k=\mathbf{P}_k^{\ell}}=\\
		&\left( \mathbf{I}_{L_{\mathrm{t}}}+ \widetilde{\mathbf{\Gamma}}_k^+ \mathbf{G}_k \left( \mathbf{t}^{\ell} \right) \mathbf{P}^{\ell} \left(\mathbf{P}^{\ell}\right)^H \mathbf{G}_k\left( \mathbf{t}^{\ell} \right)^H \right)^{-1} \widetilde{\mathbf{\Gamma}}_k^+ \mathbf{G}_k \left( \mathbf{t}^{\ell} \right) \mathbf{P}_k^{\ell}, \\
	\end{aligned}
\end{equation}
as well as
\begin{equation}
	\begin{aligned}
		\label{eq: DE Btl}
		& \left.\frac{\partial 	\bar{R}_{\mathrm{t},l}\left(\mathbf{t},\mathbf{P}\right)}{\partial \left(\mathbf{G}_l \left( \mathbf{t}\right)\mathbf{P}_k\right)^H}\right|_{\mathbf{t}=\mathbf{t}^{\ell},\mathbf{r}_k=\mathbf{r}_k^{\ell},\mathbf{P}_k=\mathbf{P}_k^{\ell}}=-\overline{\mathbf{B}}_l^{\ell} \mathbf{G}_l \left( \mathbf{t}^{\ell} \right) \mathbf{P}_k^{\ell}, \forall l\neq k,
	\end{aligned}
\end{equation}
where $\overline{\mathbf{B}}_l^{\ell}$ is defined as
\begin{equation}
	\begin{aligned}
		\overline{\mathbf{B}}_l^{\ell} &= \left( \mathbf{I}_{L_{\mathrm{t}}} + \widetilde{\mathbf{\Gamma}}_l^- \mathbf{G}_l \left( \mathbf{t}^{\ell} \right) \mathbf{P}_{\backslash l}^{\ell} \left(\mathbf{P}_{\backslash l}^{\ell}\right)^H \mathbf{G}_l^H \left( \mathbf{t}^{\ell} \right) \right)^{-1} \widetilde{\mathbf{\Gamma}}_l^- \\
		&- \left( \mathbf{I}_{L_{\mathrm{t}}}+ \widetilde{\mathbf{\Gamma}}_l^+ \mathbf{G}_l \left( \mathbf{t}^{\ell} \right) \mathbf{P}^{\ell} \left(\mathbf{P}^{\ell}\right)^H \mathbf{G}_l^H \left( \mathbf{t}^{\ell} \right) \right)^{-1} \widetilde{\mathbf{\Gamma}}_l^+.
	\end{aligned}
\end{equation}
Further, we calculate $\overline{\mathbf{A}}_k^{\ell}$ as
\begin{equation}
	\label{eq: DE Atk}
	\begin{aligned}
		\overline{\mathbf{A}}_k^{\ell} =& \left.\frac{\partial 	\bar{R}_{\mathrm{t},k}\left(\mathbf{t},\mathbf{P}\right)}{\partial \left(\mathbf{G}_k \left( \mathbf{t}\right)\mathbf{P}_k\right)^H}\right|_{\mathbf{t}=\mathbf{t}^{\ell},\mathbf{r}_k=\mathbf{r}_k^{\ell},\mathbf{P}_k=\mathbf{P}_k^{\ell}}+\overline{\mathbf{B}}_k^{\ell}\mathbf{G}_k \left( \mathbf{t}^{\ell} \right) \mathbf{P}_k^{\ell} \\
		=& \left(\mathbf{I}_{L_{\mathrm{t}}}+\widetilde{\mathbf{\Gamma}}_k^-\mathbf{G}_k \left( \mathbf{t}^{\ell} \right)\mathbf{P}_{\backslash k}^{\ell}\left(\mathbf{P}_{\backslash k}^{\ell}\right)^H\mathbf{G}_k\left( \mathbf{t}^{\ell} \right)^H \right)^{-1} \\
		&\widetilde{\mathbf{\Gamma}}_k^-\mathbf{G}_k \left( \mathbf{t}^{\ell} \right) \mathbf{P}_k^{\ell}.
	\end{aligned}
\end{equation}
Hence, comparing \eqref{eq: Btl}, \eqref{eq: Atk} and \eqref{eq: DE Btl}, \eqref{eq: DE Atk}, $\overline{\mathbf{B}}_k^{\ell}$ and $\overline{\mathbf{A}}_k^{\ell}$ are the corresponding DEs of $\mathbf{B}_k^{\ell}$ and $\mathbf{A}_k^{\ell}$, respectively. Besides, from \eqref{eq: ctk}, the DEs of $c_k^{\ell}$ can be calculated as
\begin{equation}
	\begin{aligned}
			\bar{c}_k^{\ell} &= \bar{R}_{\mathrm{t},k}(\mathbf{t}^{\ell},\mathbf{P}^{\ell}) +\operatorname{tr}\left(\overline{\mathbf{B}}_k^{\ell} \mathbf{G}_k \left( \mathbf{t}^{\ell} \right) \mathbf{P}^{\ell} \left(\mathbf{P}^{\ell}\right)^H \mathbf{G}_k\left( \mathbf{t}^{\ell} \right)^H \right) \\&- 	\operatorname{tr}\left(\left(\overline{\mathbf{A}}_k^{\ell}\right)^H \mathbf{G}_k \left( \mathbf{t}^{\ell} \right)\mathbf{P}_k^{\ell}\right) -	\operatorname{tr}\left(\overline{\mathbf{A}}_k^{\ell} \left(\mathbf{P}_k^{\ell}\right)^H\mathbf{G}_k\left( \mathbf{t}^{\ell} \right)^H \right).
	\end{aligned}
\end{equation}
Therefore, the deterministic minorizing function of the average achievable rate w.r.t. the transmit variables is obtained by replacing $\mathbf{A}_k^{\ell}$, $\mathbf{B}_k^{\ell}$ and $c_k^{\ell}$ in \eqref{eq: minorizing function t} with $\overline{\mathbf{A}}_k^{\ell}$, $\overline{\mathbf{B}}_k^{\ell}$ and $\bar{c}_k^{\ell}$:
\begin{equation}
	\label{eq: DE minorizing function t}
	\begin{aligned}
		\hat{R}_{\mathrm{t},k}^{\ell+1}\left(\mathbf{t},\mathbf{P}\right) \triangleq & \operatorname{tr}\left(\left(\overline{\mathbf{A}}_k^{\ell}\right)^H \mathbf{G}_k \left( \mathbf{t} \right)\mathbf{P}_k\right) + 	\operatorname{tr}\left(\overline{\mathbf{A}}_k^{\ell} \mathbf{P}_k^H\mathbf{G}_k\left( \mathbf{t} \right)^H \right) \\
		-& \operatorname{tr}\left(\overline{\mathbf{B}}_k^{\ell} \mathbf{G}_k \left( \mathbf{t} \right) \mathbf{P} \mathbf{P}^H \mathbf{G}_k\left( \mathbf{t} \right)^H \right) + \bar{c}_k^{\ell}.
	\end{aligned}
\end{equation}

On the other hand, the deterministic function of average achievable rate $R_k\left(\mathbf{t},\mathbf{r}_k,\mathbf{P}\right)$ w.r.t. the receive APV is given as 
\begin{equation}
	\label{eq: DE Rrk}
	\begin{aligned}
		\bar{R}_{\mathrm{r},k}\left(\mathbf{r}_k\right) \triangleq& \bar{R}_{\mathrm{r},k}^+\left(\mathbf{r}_k\right) - \bar{R}_{\mathrm{r},k}^-\left(\mathbf{r}_k\right),
	\end{aligned}
\end{equation}
where the deterministic functions are given by
\begin{equation}
	\begin{aligned}
		\bar{R}_{\mathrm{r},k}^+\left(\mathbf{r}_k\right) =& \log \operatorname{det}\left( \mathbf{I}_M + \mathbf{F}_k\left( \mathbf{r}_k \right)^H  \mathbf{\Gamma}_k^+ \mathbf{F}_k \left( \mathbf{r}_k \right) \right)
		\\+& \log \operatorname{det}\left(\mathbf{\Phi}_k^+\right) - \operatorname{tr}\left( \sigma^2 \left( \mathbf{I}_{N} - \mathbf{\Phi}_k^+ \right) \mathbf{\Theta}_k^+ \right), \\
		\bar{R}_{\mathrm{r},k}^-\left(\mathbf{r}_k\right) =& \log \operatorname{det}\left( \mathbf{I}_M + \mathbf{F}_k\left( \mathbf{r}_k \right)^H  \mathbf{\Gamma}_k^- \mathbf{F}_k \left( \mathbf{r}_k \right) \right)
		\\+& \log \operatorname{det}\left(\mathbf{\Phi}_k^-\right) - \operatorname{tr}\left( \sigma^2 \left( \mathbf{I}_{N} - \mathbf{\Phi}_k^- \right) \mathbf{\Theta}_k^- \right),
	\end{aligned}
\end{equation}
with matrices
\begin{equation}
	\begin{aligned}
		& \mathbf{\Gamma}_k^+ = - \widetilde{\eta}_k^+ \left( \mathbf{\Theta}_k^+ \right) + \sigma^{-2} \overline{\mathbf{\Sigma}}_k \mathbf{G}_k \left( \mathbf{t} \right) \mathbf{Q}^{\frac{1}{2}} \mathbf{\Phi}_k^+ \mathbf{Q}^{\frac{H}{2}} \mathbf{G}_k\left( \mathbf{t} \right)^H  \overline{\mathbf{\Sigma}}_k^H, \\
		& \mathbf{\Gamma}_k^- = - \widetilde{\eta}_k^- \left( \mathbf{\Theta}_k^- \right) + \sigma^{-2} \overline{\mathbf{\Sigma}}_k \mathbf{G}_k \left( \mathbf{t} \right) \mathbf{Q}_{\backslash k}^{\frac{1}{2}} \mathbf{\Phi}_k^- \mathbf{Q}_{\backslash k}^{\frac{H}{2}} \mathbf{G}_k\left( \mathbf{t} \right)^H  \overline{\mathbf{\Sigma}}_k^H.
	\end{aligned}
\end{equation}

\subsection{Transmit Variable Design}
\label{Section: Transmit variable Design}
After formulating $\hat{R}_{\mathrm{t},k}^{\ell}\left(\mathbf{t},\mathbf{P}\right)$, $\forall k$, we solve problem \eqref{op: EE t P} in iteration $\ell$ of the AO algorithm. While a solution to problem \eqref{op: EE t P} can be acquired by alternatively optimizing the transmit APV and the precoding matrices, it is noted that the optimization space of the transmit APV may be extremely narrowed after the optimal precoding matrices for given transmit APV are obtained. To tackle this issue, we construct an equivalent problem w.r.t. transmit APV. 

In particular, given $\mathbf{t}$, we define optimal precoding matrix $\mathbf{P}^{\ell}\left(\mathbf{t}\right)$ as the solution to the inner-layer system EE optimization problem w.r.t. the precoding matrices, i.e.,
\begin{equation}
	\label{op: EE P}
	\begin{aligned}
		\mathbf{P}^{\ell}\left(\mathbf{t}\right)\triangleq\arg \max_{\mathbf{P}} \quad & EE^{\ell}\left(\mathbf{t},\mathbf{P}\right) \\
		\text { s.t. } \quad
		& \text{C5}.
	\end{aligned}
\end{equation}
Then, problem \eqref{op: EE t P} is obviously equivalent to
\begin{equation}
	\label{op: EE t}
	\begin{aligned}
		\max_{\mathbf{t}} \quad & EE^{\ell}\left(\mathbf{t},\mathbf{P}^{\ell}\left(\mathbf{t}\right)\right)  \\
		\text { s.t. } \quad &\text{C1},\text{C2}.
	\end{aligned}
\end{equation}
Since the deterministic minorizing function in \eqref{op: EE P} is concave-convex fractional w.r.t. the precoding matrices, the optimal solution can be obtained efficiently by utilizing the fractional programming, e.g., the Dinkelbach’s method \cite{Eeiw}, while fixing the transmit APV. Thus, the gradient of the deterministic minorizing function with the optimal precoding matrices in \eqref{op: EE t}, w.r.t. the transmit APV, can be calculated efficiently by leveraging the chain rule. Therefore, for problem \eqref{op: EE t}, we can develop a low-complexity solution with the SCA method.
Next, we will propose the algorithm to solve problems \eqref{op: EE P} and \eqref{op: EE t}.

\textbf{Solving Problem \eqref{op: EE P}}: While the problem in \eqref{op: EE P} can be optimally solved by the Dinkelbach’s algorithm, the resulting subproblem in each iteration of the Dinkelbach’s algorithm still needs to tackle the power constraint. Consequently, a more efficient and well-structured iterative algorithm for the problem in \eqref{op: EE P} is developed as follows.

Different from the sum rate maximization problem, employing the full power budget for transmission may not yield optimal results for the system EE maximization design. This is primarily attributed to that the system EE tends to saturate upon the ingestion of excessive power. As a result, the pursuit of optimal transmit power consumption becomes imperative for enhancing EE performance. To explicate the relationship between the system EE and the transmit power, we commence by introducing an auxiliary function
\begin{equation}
	\label{eq: EE Pt}
	\begin{aligned}
		\widetilde{EE}^{\ell}\left(P_\mathrm{t}\right)\triangleq\max_{\mathbf{P}} \quad & EE^{\ell}\left(\mathbf{t},\mathbf{P}\right) \\
		\text { s.t. } \quad
		& \operatorname{tr}\left(\mathbf{P}\mathbf{P}^H\right) = P_\mathrm{t},
	\end{aligned}
\end{equation}
where $P_\mathrm{t}$ is an auxiliary power variable. Given transmit power $P_\mathrm{t}$, $\widetilde{EE}^{\ell}\left(P_\mathrm{t}\right)$ is the corresponding maximum system EE. Thus, the problem in \eqref{op: EE P} is equivalent to
\begin{equation}
	\begin{aligned}
		\max_{P_\mathrm{t}} \quad & \widetilde{EE}^{\ell}\left(P_\mathrm{t}\right) \\
		\text { s.t. } \quad
		& 0\leq P_\mathrm{t}\leq P_\mathrm{max}.
	\end{aligned}
\end{equation}
Note that $\widetilde{EE}^{\ell}\left(P_\mathrm{t}\right)$ is pseudo-concave w.r.t. $P_\mathrm{t}$ while fixing $\mathbf{t}$ and there exists an unique, global-optimal point \cite{Affe}. Thus, either $\widetilde{EE}^{\ell}\left(P_\mathrm{t}\right)$ is
nondecreasing in $\left[0,P_\mathrm{max}\right]$, or there exits a point $P_\mathrm{t,opt}^{\ell} \in \left[0,P_\mathrm{max}\right]$ that maximizes $\widetilde{EE}^{\ell}\left(P_\mathrm{t}\right)$ such that $\widetilde{EE}^{\ell}\left(P_\mathrm{t}\right)$ is monotonically nondecreasing when $P_\mathrm{t}\leq P_\mathrm{t,opt}^{\ell}$, and monotonically nonincreasing when $P_\mathrm{t}\geq P_\mathrm{t,opt}^{\ell}$.

Based on the above discussion, we first optimally solving the following unconstrained EE maximization problem, which is similar to the problem in \eqref{op: EE P}:
\begin{equation}
	\label{op: Unconstrained EE P}
	\begin{aligned}
		\mathbf{P}_{\mathrm{opt}}^{\ell}=\arg \max_{\mathbf{P}} \quad & EE^{\ell}\left(\mathbf{t},\mathbf{P}\right).
	\end{aligned}
\end{equation}
Then, the optimal precoding matrix for the problem in \eqref{op: EE P} equals $\mathbf{P}_{\mathrm{opt}}^{\ell}$ if the corresponding power consumption satisfies the constraint, i.e., $\operatorname{tr}\left(\mathbf{P}_{\mathrm{opt}}^{\ell}\left(\mathbf{P}_{\mathrm{opt}}^{\ell}\right)^H\right)\leq P_\mathrm{max}$. Otherwise, the maximum EE is achieved while transmitting with full power budget $P_\mathrm{max}$. Accordingly, the sum rate maximization problem is considered as follows
\begin{equation}
	\label{op: SR P}
	\begin{aligned}
		\widetilde{\mathbf{P}}_{\mathrm{opt}}^{\ell}=\arg \max_{\mathbf{P}} \quad & \sum_{k=1}^K \hat{R}_{\mathrm{t},k}^{\ell}\left(\mathbf{t},\mathbf{P}\right) \\
		\text { s.t. } \quad
		& \operatorname{tr}\left(\mathbf{P}\mathbf{P}^H\right) = P_\mathrm{max}.
	\end{aligned}
\end{equation}
Subsequently, we will present low-complexity algorithms for the above problems \eqref{op: Unconstrained EE P} and \eqref{op: SR P}.
Since problem \eqref{op: Unconstrained EE P} is a classical concave-convex fractional programming problem, it can be optimally tackled by the Dinkelbach’s method. In particular, the optimal solution to problem \eqref{op: Unconstrained EE P} can be given by $\mathbf{P}_{\mathrm{opt}}^{\ell}=\left[\mathbf{P}_{1,\mathrm{opt}}^{\ell},\mathbf{P}_{2,\mathrm{opt}}^{\ell},\cdots,\mathbf{P}_{K,\mathrm{opt}}^{\ell}\right]$, where the precoding matrices, $\mathbf{P}_{k,\mathrm{opt}}^{\ell}$, $\forall k$, are obtained through the following iterative process \cite{Eeiw, CO}:
	\begin{subequations}
		\label{eq: Dinkelbach}
		\begin{align}
			\label{eq: Dinkelbach eta}
			\eta =& EE^{\ell}\left(\mathbf{t},\mathbf{P}\right), \\
			\label{eq: Dinkelbach P}
			\nonumber
			\mathbf{P}_k=&\arg\max_{\mathbf{P}} \sum_{k=1}^K \hat{R}_{\mathrm{t},k}^{\ell}\left(\mathbf{t},\mathbf{P}\right) - \eta P_\mathrm{tot} \\\nonumber
			=&\left( \sum_{k=1}^K\mathbf{G}_k\left( \mathbf{t} \right)^H \overline{\mathbf{B}}_k^{\ell-1}\mathbf{G}_k \left( \mathbf{t} \right) + \eta\mathbf{I}_N\right)^{-1}\\ & \mathbf{G}_k\left( \mathbf{t} \right)^H \overline{\mathbf{A}}_k^{\ell-1},\forall k.
		\end{align}
	\end{subequations}	
	At first, $\mathbf{P}$ is initialized to satisfy the power constraint C5, and then, $\eta$ and $\mathbf{P}$ are updated according to \eqref{eq: Dinkelbach eta} and \eqref{eq: Dinkelbach P}, respectively, until $\eta$ converges.
	On the other hand, the optimal solution to solve problem \eqref{op: SR P} can be given by $\widetilde{\mathbf{P}}_{\mathrm{opt}}^{\ell}=\left[\widetilde{\mathbf{P}}_{1,\mathrm{opt}}^{\ell},\widetilde{\mathbf{P}}_{2,\mathrm{opt}}^{\ell},\cdots,\widetilde{\mathbf{P}}_{K,\mathrm{opt}}^{\ell}\right]$, where the precoding matrices, $\widetilde{\mathbf{P}}_{k,\mathrm{opt}}^{\ell}$, $\forall k$, follow the classical water-filling structure as follows \cite{CO}
	\begin{equation}
		\begin{aligned}
			\label{eq: water-filling}
			\widetilde{\mathbf{P}}_{k,\mathrm{opt}}^{\ell}=&\left( \sum_{k=1}^K\mathbf{G}_k\left( \mathbf{t} \right)^H \overline{\mathbf{B}}_k^{\ell-1}\mathbf{G}_k \left( \mathbf{t} \right) + \mu\mathbf{I}_N\right)^{-1}\\ & \mathbf{G}_k\left( \mathbf{t} \right)^H\overline{\mathbf{A}}_k^{\ell-1}, \forall k,
		\end{aligned}
	\end{equation}
	with $\mu$ satisfying $\operatorname{tr}\left(\widetilde{\mathbf{P}}_{\mathrm{opt}}^{\ell}\left(\widetilde{\mathbf{P}}_{\mathrm{opt}}^{\ell}\right)^H\right) = P_\mathrm{max}$. The appropriate $\mu$ can be found by the bisection method. Additionally, matrix inversions $\left( \sum_{k=1}^K\mathbf{G}_k\left( \mathbf{t} \right)^H \overline{\mathbf{B}}_k^{\ell-1}\mathbf{G}_k \left( \mathbf{t} \right) + \eta\mathbf{I}_N\right)^{-1}$ and $\left( \sum_{k=1}^K\mathbf{G}_k\left( \mathbf{t} \right)^H \overline{\mathbf{B}}_k^{\ell-1}\mathbf{G}_k \left( \mathbf{t} \right) + \mu\mathbf{I}_N\right)^{-1}$ for different $\eta$ and $\mu$ in the iterative process of the Dinkelbach's method and the bisection method can be efficiently computed as $\mathbf{U}\left( \mathbf{\Lambda} + \eta\mathbf{I}_N\right)^{-1}\mathbf{U}^H$ and $\mathbf{U}\left( \mathbf{\Lambda} + \mu\mathbf{I}_N\right)^{-1}\mathbf{U}^H$, respectively, where $\mathbf{U}$  is the square $N\times N$ matrix whose each column is the eigenvector of matrix $\sum_{k=1}^K\mathbf{G}_k\left( \mathbf{t} \right)^H \overline{\mathbf{B}}_k^{\ell-1}\mathbf{G}_k \left( \mathbf{t} \right)$ and $\mathbf{\Lambda}$ is the diagonal matrix whose main diagonal elements are the corresponding eigenvalues.

\textbf{Solving Problem \eqref{op: EE t}}: Now, we present the SCA algorithm to solve problem \eqref{op: EE t}. To construct the convex surrogate for optimizing the transmit APV, the gradient of optimal $EE^{\ell}\left(\mathbf{t},\mathbf{P}^{\ell}\left(\mathbf{t}\right)\right)$ is derived according to the chain rule, which is given in \eqref{eq: gradient t (a)} at the bottom of the next page, where ($a$) holds since $\mathbf{P}_{\mathrm{opt}}^{\ell}$ is the optimal solution of problem \eqref{op: Unconstrained EE P}, and thus, its first-order optimality condition $\frac{\partial}{\partial\mathbf{P}} EE^{\ell}\left(\mathbf{t},\mathbf{P}_{\mathrm{opt}}^{\ell}\right)=\mathbf{0}$ is satisfied; $\frac{\partial}{\partial \mathbf{t}} EE^{\ell}\left(\mathbf{t},\mathbf{P}\right)$ and $\frac{\partial}{\partial\mathbf{P}} EE^{\ell}\left(\mathbf{t},\mathbf{P}\right)$ are given by \eqref{eq: partialt} and \eqref{eq: partialPk}, respectively; $\frac{\partial\mu}{\partial\mathbf{t}}$ is calculated from the fact that $\nabla_{\mathbf{t}}\operatorname{tr}\left(\widetilde{\mathbf{P}}_{\mathrm{opt}}^{\ell}\left(\widetilde{\mathbf{P}}_{\mathrm{opt}}^{\ell}\right)^H\right) = \frac{\partial}{\partial\mathbf{t}}\operatorname{tr}\left(\widetilde{\mathbf{P}}_{\mathrm{opt}}^{\ell}\left(\widetilde{\mathbf{P}}_{\mathrm{opt}}^{\ell}\right)^H\right) + \frac{\partial\mu}{\partial\mathbf{t}} \frac{\partial}{\partial\mu}\operatorname{tr}\left(\widetilde{\mathbf{P}}_{\mathrm{opt}}^{\ell}\left(\widetilde{\mathbf{P}}_{\mathrm{opt}}^{\ell}\right)^H\right)  = \nabla_{\mathbf{t}} P_\mathrm{max} =0$; $\left( \cdot\right)$ represents $\left( \sum_{k=1}^K\mathbf{G}_k\left( \mathbf{t} \right)^H \overline{\mathbf{B}}_k^{\ell-1}\mathbf{G}_k \left( \mathbf{t} \right) + \mu\mathbf{I}_N\right)$ for simplicity of exposition.
\begin{figure*}[!b]
	\hrule
	\begin{subequations}
		\label{eq: gradient t}
		\begin{align}
			\label{eq: gradient t (a)}
			\nonumber
			&\nabla_{\mathbf{t}} 	EE^{\ell}\left(\mathbf{t},\mathbf{P}^{\ell}\left(\mathbf{t}\right)\right) =\\&\qquad\qquad \begin{cases}	
				\frac{\partial}{\partial \mathbf{t}} EE^{\ell}\left(\mathbf{t},\mathbf{P}_{\mathrm{opt}}^{\ell}\right)
				+2\operatorname{Re}\left\{\left(\frac{1}{\partial\mathbf{t}}+\nabla_{\mathbf{t}}\eta\frac{1}{\partial\eta}\right) \operatorname{tr}\left(\left( \frac{\partial}{\partial\mathbf{P}} EE^{\ell}\left(\mathbf{t},\mathbf{P}_{\mathrm{opt}}^{\ell}\right)\right)^H \partial\mathbf{P}_{\mathrm{opt}}^{\ell}\right)\right\} \\\qquad
				\overset{(a)}{=}\frac{\partial}{\partial \mathbf{t}} EE^{\ell}\left(\mathbf{t},\mathbf{P}_{\mathrm{opt}}^{\ell}\right), \qquad\qquad\qquad\qquad\qquad\qquad\qquad\qquad \text{ if } \operatorname{tr}\left(\mathbf{P}_{\mathrm{opt}}^{\ell}\left(\mathbf{P}_{\mathrm{opt}}^{\ell}\right)^H\right)\leq P_\mathrm{max}, \\
				\frac{\partial}{\partial \mathbf{t}} EE^{\ell}\left(\mathbf{t},\widetilde{\mathbf{P}}_{\mathrm{opt}}^{\ell}\right) +2\operatorname{Re}\left\{\left(\frac{1}{\partial\mathbf{t}}+\frac{\partial\mu}{\partial\mathbf{t}}\frac{1}{\partial\mu}\right) \operatorname{tr}\left(\left( \frac{\partial}{\partial\mathbf{P}} EE^{\ell}\left(\mathbf{t},\widetilde{\mathbf{P}}_{\mathrm{opt}}^{\ell}\right)\right)^H \partial\widetilde{\mathbf{P}}_{\mathrm{opt}}^{\ell}\right)\right\} \\
				\qquad
				= \frac{\partial}{\partial \mathbf{t}} EE^{\ell}\left(\mathbf{t},\widetilde{\mathbf{P}}_{\mathrm{opt}}^{\ell}\right) +2\operatorname{Re}\left\{ \sum_{k=1}^K \mathbf{\Delta}_{\mathrm{t},k}\left(\left( \frac{\partial}{\partial\mathbf{P}_k} EE^{\ell}\left(\mathbf{t},\widetilde{\mathbf{P}}_{\mathrm{opt}}^{\ell}\right)\right)^H\right) \right.\\\left.
				\qquad- \frac{ \left(\sum_{k=1}^K \operatorname{Re}\left\{ \mathbf{\Delta}_{\mathrm{t},k}\left(\left(\widetilde{\mathbf{P}}_{k,\mathrm{opt}}^{\ell}\right)^H\right) \right\}\right) \left(\sum_{k=1}^K \operatorname{tr}\left(\left( \frac{\partial}{\partial\mathbf{P}_k} EE^{\ell}\left(\mathbf{t},\widetilde{\mathbf{P}}_{\mathrm{opt}}^{\ell}\right)\right)^H
					\left( \cdot\right)^{-2}\mathbf{G}_k\left( \mathbf{t} \right)^H\overline{\mathbf{A}}_k^{\ell-1}\right) \right)}{\sum_{k=1}^K\operatorname{Re}\left\{ \operatorname{tr}\left( \left(\widetilde{\mathbf{P}}_{k,\mathrm{opt}}^{\ell}\right)^H
				\left( \cdot\right)^{-2}\mathbf{G}_k\left( \mathbf{t} \right)^H\overline{\mathbf{A}}_k^{\ell-1}\right)\right\}} \right\},\text{ otherwise};
			\end{cases}\\
			\nonumber
			&\frac{\partial}{\partial \mathbf{t}} EE^{\ell}\left(\mathbf{t},\mathbf{P}\right) =\frac{2}{P_\mathrm{tot}} \sum_{k=1}^K \operatorname{Re}\left\{
			\begin{gathered}
				\begin{bmatrix} 
					\operatorname{diag}\left\{\mathbf{P}_k\left(\overline{\mathbf{A}}_k^{\ell-1}\right)^H \mathbf{\Delta}_{k,x_{\mathrm{t}}}\mathbf{G}_k \left( \mathbf{t} \right)\right\} \\
					\operatorname{diag}\left\{\mathbf{P}_k\left(\overline{\mathbf{A}}_k^{\ell-1}\right)^H \mathbf{\Delta}_{k,y_{\mathrm{t}}}\mathbf{G}_k \left( \mathbf{t} \right)\right\}
				\end{bmatrix}
			\end{gathered} -\begin{gathered}
				\begin{bmatrix} 
					\operatorname{diag}\left\{\mathbf{P} \mathbf{P}^H \mathbf{G}_k\left( \mathbf{t} \right)^H \overline{\mathbf{B}}_k^{\ell-1} \mathbf{\Delta}_{k,x_{\mathrm{t}}}\mathbf{G}_k \left( \mathbf{t} \right)\right\} \\
					\operatorname{diag}\left\{\mathbf{P} \mathbf{P}^H \mathbf{G}_k\left( \mathbf{t} \right)^H \overline{\mathbf{B}}_k^{\ell-1} \mathbf{\Delta}_{k,y_{\mathrm{t}}}\mathbf{G}_k \left( \mathbf{t} \right)\right\}
				\end{bmatrix}
			\end{gathered}\right\},\\
			\label{eq: partialt}
			&\mathbf{\Delta}_{k,x_{\mathrm{t}}} \triangleq \operatorname{Diag}\left\{ \frac{j2\pi}{\lambda} \left( \sin\theta_{\mathrm{t},k}^{1}\cos\phi_{\mathrm{t},k}^{1},\cdots,\sin\theta_{\mathrm{t},k}^{L_{\mathrm{t}}}\cos\phi_{\mathrm{t},k}^{L_{\mathrm{t}}} \right)^T\right\},
			\mathbf{\Delta}_{k,y_{\mathrm{t}}} \triangleq \operatorname{Diag}\left\{ \frac{j2\pi}{\lambda} \left(
			\cos\theta_{\mathrm{t},k}^{1},\cdots,\cos\theta_{\mathrm{t},k}^{L_{\mathrm{t}}} \right)^T\right\};\\
			\label{eq: partialPk}
			&\frac{\partial}{\partial\mathbf{P}_k} EE^{\ell}\left(\mathbf{t},\mathbf{P}\right)= \frac{1}{P_\mathrm{tot}} \left( \mathbf{G}_k \left( \mathbf{t} \right)^H \overline{\mathbf{A}}_k^{\ell-1}
			- \left( \sum_{k=1}^K \mathbf{G}_k\left( \mathbf{t} \right)^H\overline{\mathbf{B}}_k^{\ell-1} \mathbf{G}_k \left( \mathbf{t} \right) \right) \mathbf{P}_k \right);\\
			\nonumber
			&\mathbf{\Delta}_{\mathrm{t},k}\left(\mathbf{Y}\right) \triangleq 
			\begin{gathered}
				\begin{bmatrix} 
					\operatorname{diag}\left\{ \mathbf{G}_k\left( \mathbf{t} \right)^H\mathbf{\Delta}_{k,x_{\mathrm{t}}}^H\overline{\mathbf{A}}_k^{\ell-1} \mathbf{Y} \left( \cdot\right)^{-1} \right\} \\
					\operatorname{diag}\left\{ \mathbf{G}_k\left( \mathbf{t} \right)^H\mathbf{\Delta}_{k,y_{\mathrm{t}}}^H\overline{\mathbf{A}}_k^{\ell-1} \mathbf{Y} \left( \cdot\right)^{-1} \right\}
				\end{bmatrix}
			\end{gathered} -
			\begin{gathered}
				\begin{bmatrix} 
					\operatorname{diag}\left\{ \left(\sum_{k=1}^K\mathbf{G}_k\left( \mathbf{t} \right)^H\mathbf{\Delta}_{k,x_{\mathrm{t}}}^H \overline{\mathbf{B}}_k^{\ell-1}\mathbf{G}_k \left( \mathbf{t} \right) \right)\left( \cdot\right)^{-1}\mathbf{G}_k\left( \mathbf{t} \right)^H\overline{\mathbf{A}}_k^{\ell-1} \mathbf{Y} \left( \cdot\right)^{-1} \right\} \\
					\operatorname{diag}\left\{ \left(\sum_{k=1}^K\mathbf{G}_k\left( \mathbf{t} \right)^H\mathbf{\Delta}_{k,y_{\mathrm{t}}}^H \overline{\mathbf{B}}_k^{\ell-1}\mathbf{G}_k \left( \mathbf{t} \right) \right)\left( \cdot\right)^{-1}\mathbf{G}_k\left( \mathbf{t} \right)^H\overline{\mathbf{A}}_k^{\ell-1} \mathbf{Y} \left( \cdot\right)^{-1} \right\}
				\end{bmatrix}
			\end{gathered} \\
			\label{eq: Deltatk}
			&\qquad\qquad\qquad-
			\begin{gathered}
				\begin{bmatrix} 
					\operatorname{diag}\left\{ \left( \cdot\right)^{-1}\mathbf{G}_k\left( \mathbf{t} \right)^H\overline{\mathbf{A}}_k^{\ell-1} \mathbf{Y} \left( \cdot\right)^{-1} \left(\sum_{k=1}^K\mathbf{G}_k\left( \mathbf{t} \right)^H \overline{\mathbf{B}}_k^{\ell-1} \mathbf{\Delta}_{k,x_{\mathrm{t}}}\mathbf{G}_k\left( \mathbf{t} \right)\right)  \right\} \\
					\operatorname{diag}\left\{ \left( \cdot\right)^{-1}\mathbf{G}_k\left( \mathbf{t} \right)^H\overline{\mathbf{A}}_k^{\ell-1} \mathbf{Y} \left( \cdot\right)^{-1} \left(\sum_{k=1}^K\mathbf{G}_k\left( \mathbf{t} \right)^H \overline{\mathbf{B}}_k^{\ell-1} \mathbf{\Delta}_{k,y_{\mathrm{t}}}\mathbf{G}_k\left( \mathbf{t} \right)\right) \right\}
				\end{bmatrix}
			\end{gathered}.
		\end{align}
	\end{subequations}
\end{figure*}
Then, given $\mathbf{t}^{\ell,\hat{\ell}_{\mathrm{t}}}$, the concave surrogate of $EE^{\ell}\left(\mathbf{t},\mathbf{P}^{\ell}\left(\mathbf{t}\right)\right)$ w.r.t. $\mathbf{t}$ can be formulated as
\begin{equation}
	\begin{aligned}
		\widehat{EE}^{\ell,\hat{\ell}_{\mathrm{t}}+1}\left(\mathbf{t}\right) &\triangleq -\delta_{\mathrm{t}}\left(\mathbf{t}-\mathbf{t}^{\ell,\hat{\ell}_{\mathrm{t}}}\right)^T\left(\mathbf{t}-\mathbf{t}^{\ell,\hat{\ell}_{\mathrm{t}}}\right) \\ &+ \nabla_{\mathbf{t}} 	EE^{\ell}\left(\mathbf{t}^{\ell,\hat{\ell}_{\mathrm{t}}},\mathbf{P}^{\ell}\left(\mathbf{t}^{\ell,\hat{\ell}_{\mathrm{t}}}\right)\right)^T \left(\mathbf{t}-\mathbf{t}^{\ell,\hat{\ell}_{\mathrm{t}}}\right),
	\end{aligned}
\end{equation}
where $\delta_{\mathrm{t}}>0$ is a positive number to guarantee the strong concavity of the objective. On the other hand, given $\mathbf{t}^{\ell,\hat{\ell}_{\mathrm{t}}}$, we have $\left\|\mathbf{t}_i-\mathbf{t}_j\right\|_2 \geq \frac{1}{\left\|\mathbf{t}_i^{\ell,\hat{\ell}_{\mathrm{t}}}-\mathbf{t}_j^{\ell,\hat{\ell}_{\mathrm{t}}}\right\|_2}\left(\mathbf{t}_i-\mathbf{t}_j\right)^T\left(\mathbf{t}_i^{\ell,\hat{\ell}_{\mathrm{t}}}-\mathbf{t}_j^{\ell,\hat{\ell}_{\mathrm{t}}}\right)$, $\forall i \neq j$. That is, the position constraints of the transmit MAs in C2 can be replaced with
\begin{equation}
	\label{eq: t constraint surrogate}
	\begin{aligned}
		\frac{1}{\left\|\mathbf{t}_i^{\ell,\hat{\ell}_{\mathrm{t}}}-\mathbf{t}_j^{\ell,\hat{\ell}_{\mathrm{t}}}\right\|_2}\left(\mathbf{t}_i-\mathbf{t}_j\right)^T\left(\mathbf{t}_i^{\ell,\hat{\ell}_{\mathrm{t}}}-\mathbf{t}_j^{\ell,\hat{\ell}_{\mathrm{t}}}\right) \geq D,\quad \forall i \neq j.
	\end{aligned}
\end{equation}

Accordingly, the convex surrogate in iteration $\hat{\ell}_{\mathrm{t}}+1$ of the proposed SCA algorithm for problem \eqref{op: EE t} can be given by
\begin{equation}
	\label{op: SCA t surrogate}
	\begin{aligned}
		\max_{\mathbf{t}} \quad & \widehat{EE}^{\ell,\hat{\ell}_{\mathrm{t}}+1}\left(\mathbf{t}\right) \\
		\text { s.t. } \quad
		&\text{C1},\eqref{eq: t constraint surrogate}.
	\end{aligned}
\end{equation}
To reduce computational complexity, we consider the problem similar to that in \eqref{op: SCA t surrogate} but without constraint \eqref{eq: t constraint surrogate}, whose solution is trivially 
\begin{equation}
	\label{eq: temp optimal t}
	\begin{aligned}
		\tilde{\mathbf{t}}^{\ell,\hat{\ell}_{\mathrm{t}}+1}=\left[ \mathbf{t}^{\ell,\hat{\ell}_{\mathrm{t}}}+\frac{\nabla_{\mathbf{t}} 	EE^{\ell}\left(\mathbf{t}^{\ell,\hat{\ell}_{\mathrm{t}}},\mathbf{P}^{\ell}\left(\mathbf{t}^{\ell,\hat{\ell}_{\mathrm{t}}}\right)\right)}{2\delta_{\mathrm{t}}} \right]_{\mathcal{C}_{\mathrm{t}}},
	\end{aligned}
\end{equation}
where $\left[ \mathbf{x} \right]_{\mathcal{X}}$ projects $\mathbf{x}$ into region $\mathcal{X}$.
If $\tilde{\mathbf{t}}^{\ell,\hat{\ell}_{\mathrm{t}}+1}$ satisfies constraint \eqref{eq: t constraint surrogate}, it is also the optimal solution to problem \eqref{op: SCA t surrogate}. Otherwise, problem \eqref{op: SCA t surrogate} can be efficiently solved by traditional convex optimization techniques, such as CVX. Then, the transmit APV is updated according to
\begin{equation}
	\label{eq: update t}
	\begin{aligned}
		\mathbf{t}^{\ell,\hat{\ell}_{\mathrm{t}}+1}=\mathbf{t}^{\ell,\hat{\ell}_{\mathrm{t}}}+\tau_{\mathrm{t}}^{\ell,\hat{\ell}_{\mathrm{t}}+1} \left(\bar{\mathbf{t}}^{\ell,\hat{\ell}_{\mathrm{t}}+1}-\mathbf{t}^{\ell,\hat{\ell}_{\mathrm{t}}}\right),
	\end{aligned}
\end{equation}
where $\bar{\mathbf{t}}^{\ell,\hat{\ell}_{\mathrm{t}}+1}$ and $\tau_{\mathrm{t}}^{\ell,\hat{\ell}_{\mathrm{t}}+1}$ are the solution to problem \eqref{op: SCA t surrogate} and the step size selected by backtracking line search, respectively. In each iteration, we start with a large positive step size, $\tau_{\mathrm{t}}^{\ell,\hat{\ell}_{\mathrm{t}}+1}=\tau^0\in (0, 1]$, and repeatedly reduce it to $\kappa\tau_{\mathrm{t}}^{\ell,\hat{\ell}_{\mathrm{t}}+1}$ with factor $\kappa\in(0,1)$, until the following Armijo–Goldstein condition is satisfied:
\begin{equation}
	\label{eq: Armijo–Goldstein condition t}
	\begin{aligned}
		I_{\mathrm{t}}^{\ell}\left(\mathbf{t}^{\ell,\hat{\ell}_{\mathrm{t}}+1},\mathbf{t}^{\ell,\hat{\ell}_{\mathrm{t}}}\right) \geq \xi \tau_{\mathrm{t}}^{\ell,\hat{\ell}_{\mathrm{t}}+1} \left\|	\bar{\mathbf{t}}^{\ell,\hat{\ell}_{\mathrm{t}}+1}-\mathbf{t}^{\ell,\hat{\ell}_{\mathrm{t}}} \right\|^2,
	\end{aligned}
\end{equation}
where $I_{\mathrm{t}}^{\ell}\left(\mathbf{t}^{\ell,\hat{\ell}_{\mathrm{t}}+1},\mathbf{t}^{\ell,\hat{\ell}_{\mathrm{t}}}\right) \triangleq EE^{\ell}\left(\mathbf{t}^{\ell,\hat{\ell}_{\mathrm{t}}+1},\mathbf{P}^{\ell}\left(\mathbf{t}^{\ell,\hat{\ell}_{\mathrm{t}}+1}\right)\right) - EE^{\ell}\left(\mathbf{t}^{\ell,\hat{\ell}_{\mathrm{t}}},\mathbf{P}^{\ell}\left(\mathbf{t}^{\ell,\hat{\ell}_{\mathrm{t}}}\right)\right)$ is the increment of the objective in problem \eqref{op: EE t} and $\xi \in (0, 1)$ is a given control parameter to guarantee that  the objective function achieves an adequate increase. Note that a smaller $\delta_{\mathrm{t}}$ typically leads to a more appropriate update $\tau_{\mathrm{t}}^{\ell,\hat{\ell}_{\mathrm{t}}+1} \left(\bar{\mathbf{t}}^{\ell,\hat{\ell}_{\mathrm{t}}+1}-\mathbf{t}^{\ell,\hat{\ell}_{\mathrm{t}}}\right)$ in \eqref{eq: update t} since a potential update  $\left(\bar{\mathbf{t}}^{\ell,\hat{\ell}_{\mathrm{t}}+1}-\mathbf{t}^{\ell,\hat{\ell}_{\mathrm{t}}}\right)$ is larger and $\tau_{\mathrm{t}}^{\ell,\hat{\ell}_{\mathrm{t}}+1}$ can be properly selected by backtracking line search, while it tends to cause $\tilde{\mathbf{t}}^{\ell,\hat{\ell}_{\mathrm{t}}+1}$ obtained by \eqref{eq: temp optimal t} to violate constraint \eqref{eq: t constraint surrogate}, introducing the additional computational complexity for directly solving problem \eqref{op: SCA t surrogate}. The SCA algorithm is terminated when $\hat{\ell}_{\mathrm{t}} = \hat{L}_{\mathrm{t}}$ or the increment is less than $\epsilon_1$, i.e.,
\begin{equation}
	\label{eq: termin condition t}
	\begin{aligned}
		I_{\mathrm{t}}^{\ell}\left(\mathbf{t}^{\ell,\hat{\ell}_{\mathrm{t}}+1},\mathbf{t}^{\ell,\hat{\ell}_{\mathrm{t}}}\right) \leq \epsilon_1.
	\end{aligned}
\end{equation}

\begin{algorithm}[!h]
	\caption{Algorithm for Solving Problem \eqref{op: EE t P}}
	\label{Al: SCA t}
	\begin{spacing}{0.9}
		\KwIn{$N$, $K$, $\lambda$, $D$, $\mathcal{C}_{\mathrm{t}}$, $\omega$, $P_{\mathrm{max}}$, $P_{\mathrm{c}}$, $P_{\mathrm{s}}$; $\delta_{\mathrm{t}}$, $\epsilon_1$, $\hat{L}_{\mathrm{t}}$, $\tau^0$, $\kappa$, $\xi$}
		\KwOut{$\mathbf{t}^{\ell}$ and $\mathbf{P}^{\ell}$}
		
		Initialize $\hat{\ell}_{\mathrm{t}}=0$ and $\mathbf{t}^{\ell,0}=\mathbf{t}^{\ell-1}$\
		
		\Repeat{\eqref{eq: termin condition t} is satisfied and $\hat{\ell}_{\mathrm{t}} = \hat{L}_{\mathrm{t}}$}{
			
			Calculate $\nabla_{\mathbf{t}} 	EE^{\ell}\left(\mathbf{t}^{\ell,\hat{\ell}_{\mathrm{t}}},\mathbf{P}^{\ell}\left(\mathbf{t}^{\ell,\hat{\ell}_{\mathrm{t}}}\right)\right)$ by \eqref{eq: gradient t} and $\tilde{\mathbf{t}}^{\ell,\hat{\ell}_{\mathrm{t}}+1}$ by \eqref{eq: temp optimal t}\
			
			\eIf{$\tilde{\mathbf{t}}^{\ell,\hat{\ell}_{\mathrm{t}}+1}$ satisfyies constraint \eqref{eq: t constraint surrogate}}
			{Set $\bar{\mathbf{t}}^{\ell,\hat{\ell}_{\mathrm{t}}+1}=\tilde{\mathbf{t}}^{\ell,\hat{\ell}_{\mathrm{t}}+1}$}
			{Calculate $\bar{\mathbf{t}}^{\ell,\hat{\ell}_{\mathrm{t}}+1}$ by solving problem \eqref{op: SCA t surrogate}}
			
			Initialize $\tau_{\mathrm{t}}^{\ell,\hat{\ell}_{\mathrm{t}}+1}=\tau^0/\kappa$\
			
			\Repeat{\eqref{eq: Armijo–Goldstein condition t} is satisfied}
			{Set $\tau_{\mathrm{t}}^{\ell,\hat{\ell}_{\mathrm{t}}+1}=\kappa\tau_{\mathrm{t}}^{\ell,\hat{\ell}_{\mathrm{t}}+1}$\
				
				Update $\mathbf{t}^{\ell,\hat{\ell}_{\mathrm{t}}+1}$ by \eqref{eq: update t}}
			
			Set $\hat{\ell}_{\mathrm{t}} = \hat{\ell}_{\mathrm{t}}+1$\
			
		}
		
		Calculate $\mathbf{P}^{\ell}$ by solving problem \eqref{op: EE P}\
		
		\textbf{return} $\mathbf{t}^{\ell}=\mathbf{t}^{\ell,\hat{L}_{\mathrm{t}}}$ and $\mathbf{P}^{\ell}$\
	\end{spacing}
\end{algorithm}
The overall SCA algorithm is summarized in Algorithm \ref{Al: SCA t}. In particular, $\mathbf{t}^{\ell,0}$ is initialized as the output of the AO algorithm in the last iteration (i.e., $\mathbf{t}^{\ell-1}$) in Step 1 and then optimized by the SCA algorithm in Steps 2-15. In Steps 3-8, problem \eqref{op: SCA t surrogate} is constructed and solved to obtain solution $\bar{\mathbf{t}}^{\ell,\hat{\ell}_{\mathrm{t}}+1}$. In Steps 9-13, the transmit APV $\mathbf{t}^{\ell,\hat{\ell}_{\mathrm{t}}+1}$ is updated with proper step size $\tau_{\mathrm{t}}^{\ell,\hat{\ell}_{\mathrm{t}}+1}$ chosen by backtracking line search. Finally, the optimal precoding matrix $\mathbf{P}^{\ell}$ is achieved by solving problem \eqref{op: EE P} in Step 16. The convergence of Algorithm \ref{Al: SCA t} is analyzed as follows. 
The solution of problem \eqref{op: SCA t surrogate} gives a feasible ascent direction of problem \eqref{op: EE t} at $\mathbf{t}^{\ell,\hat{\ell}_{\mathrm{t}}}$, i.e., $\left(\bar{\mathbf{t}}^{\ell,\hat{\ell}_{\mathrm{t}}+1}-\mathbf{t}^{\ell,\hat{\ell}_{\mathrm{t}}}\right)$ \cite{DbPL}. We can always find a sufficiently small positive $\tau_{\mathrm{t}}^{\ell,\hat{\ell}_{\mathrm{t}}+1}$ to ensure that the objective is increasing until $\bar{\mathbf{t}}^{\ell,\hat{\ell}_{\mathrm{t}}+1}=\mathbf{t}^{\ell,\hat{\ell}_{\mathrm{t}}}$. Besides, the objective is upper-bounded by a finite value since the feasible region is compact. Therefore, the objective function sequence convergence of Algorithm \ref{Al: SCA t} is guaranteed.
Moreover, the computational complexity of Algorithm \ref{Al: SCA t} is mainly determined by calculating $\mathbf{P}^{\ell}\left(\mathbf{t}^{\ell,\hat{\ell}_{\mathrm{t}}}\right)$ by \eqref{eq: Dinkelbach}/\eqref{eq: water-filling} and solving problem \eqref{op: SCA t surrogate} in each iteration. Specifically, the former needs computing the eigenvalue decomposition of a $N\times N$ matrix, which entails $\mathcal{O}\left(N^3\right)$ and the latter requires $\mathcal{O}\left(N^{3.5}\log(1/\epsilon)\right)$ when the interior-point method is employed with accuracy $\epsilon$. Therefore, the overall computational complexity of Algorithm \ref{Al: SCA t} is $\mathcal{O}\left(\hat{L}_{\mathrm{t}}N^{3.5}\log(1/\epsilon)\right)$.

\subsection{Receive APV Design}
\label{Section: Receive APV Design}
In this subsection, we propose a low-complexity SCA algorithm to solve problem \eqref{op: SR r} efficiently, where the structure of the objective is leveraged to construct a series of convex surrogate problems.

Define the matrix-valued function as $\mathbf{J}_k^+\left(\mathbf{r}_k\right) \triangleq \mathbf{F}_k\left( \mathbf{r}_k \right)^H  \mathbf{\Gamma}_k^+ \mathbf{F}_k \left( \mathbf{r}_k \right)$ and then the first differential and the second differential of $\bar{R}_{\mathrm{r},k}^+\left(\mathbf{r}_k\right)$ can be written as
\begin{equation}
	\begin{aligned}
		\partial \bar{R}_{\mathrm{r},k}^+\left(\mathbf{r}_k\right) &=\operatorname{tr} \left( \mathbf{E}_k^+ \left( \partial \mathbf{J}_k^+\left(\mathbf{r}_k\right) \right)\right),
	\end{aligned}
\end{equation}
\begin{equation}
	\begin{aligned}
		\partial^2 \bar{R}_{\mathrm{r},k}^+\left(\mathbf{r}_k\right) &=-\operatorname{tr} \left( \mathbf{E}_k^+ \left( \partial \mathbf{J}_k^+\left(\mathbf{r}_k\right) \right) \mathbf{E}_k^+ \left( \partial \mathbf{J}_k^+\left(\mathbf{r}_k\right) \right) \right),
	\end{aligned}
\end{equation}
respectively, where $\mathbf{E}_k^+=\left( \mathbf{I}_M + \mathbf{F}_k\left( \mathbf{r}_k \right)^H  \mathbf{\Gamma}_k^+ \mathbf{F}_k \left( \mathbf{r}_k \right) \right)^{-1}$. Since the function $\bar{R}_{\mathrm{r},k}^+\left(\mathbf{r}_k\right)$ exhibits concavity w.r.t. matrix $\mathbf{J}_k^+\left(\mathbf{r}_k\right)$, in iteration $\hat{\ell}_{\mathrm{r}}+1$ of the  proposed SCA algorithm for problem \eqref{op: SR r}, a surrogate function that maintains this concavity property can be formulated as
	\begin{equation}
		\label{eq: Rhat_Jk+}
		\begin{aligned}
			&\hat{R}_{J,k}^+\left(\mathbf{J}_k^+\left(\mathbf{r}_k\right)\right) =\operatorname{tr} \left( \mathbf{E}_k^+ 	\left( \mathbf{J}_k^+\left(\mathbf{r}_k\right)-\mathbf{J}_k^+\left(\mathbf{r}_k^{\ell,\hat{\ell}_{\mathrm{r}}}\right) \right) \right)\\&+ \bar{R}_{\mathrm{r},k}^+\left(\mathbf{r}_k^{\ell,\hat{\ell}_{\mathrm{r}}}\right)-\operatorname{tr} \left( \mathbf{E}_k^+ \left( \mathbf{J}_k^+\left(\mathbf{r}_k\right)-\mathbf{J}_k^+\left(\mathbf{r}_k^{\ell,\hat{\ell}_{\mathrm{r}}}\right) \right)\right.\\&\left.\qquad\qquad\qquad\qquad \mathbf{E}_k^+ \left( \mathbf{J}_k^+\left(\mathbf{r}_k\right)-\mathbf{J}_k^+\left(\mathbf{r}_k^{\ell,\hat{\ell}_{\mathrm{r}}}\right) \right) \right).
		\end{aligned}
	\end{equation}
The concave surrogate of $\bar{R}_{\mathrm{r},k}^+\left(\mathbf{r}_k\right)$ w.r.t $\mathbf{r}_k$ can be constructed by linearizing function $\mathbf{J}_k^+\left(\mathbf{r}_k\right)$ in $\hat{R}_{J,k}^+\left(\mathbf{J}_k^+\left(\mathbf{r}_k\right)\right)$.
In particular, we define matrices 
\begin{equation}
	\label{eq: Delta XYr}
	\begin{aligned}
		\mathbf{\Delta}_{k,X_{\mathrm{r}}} =& \mathbf{F}_k\left( \mathbf{r}_k \right)^H  \mathbf{\Gamma}_k^+ \mathbf{\Delta}_{k,x_{\mathrm{r}}} \mathbf{F}_k \left( \mathbf{r}_k \right), \\
		\mathbf{\Delta}_{k,Y_{\mathrm{r}}} =& \mathbf{F}_k\left( \mathbf{r}_k \right)^H  \mathbf{\Gamma}_k^+ \mathbf{\Delta}_{k,y_{\mathrm{r}}} \mathbf{F}_k \left( \mathbf{r}_k \right),
	\end{aligned}
\end{equation}
where
\begin{equation}
	\label{eq: Delta xyr}
	\begin{aligned}
		\mathbf{\Delta}_{k,x_{\mathrm{r}}} \triangleq& \operatorname{Diag}\left\{ \frac{j2\pi}{\lambda} \left( \sin\theta_{\mathrm{r},k}^{1}\cos\phi_{\mathrm{r},k}^{1}, \right.\right.\\&\left.\left. \quad\qquad\qquad
		\cdots,\sin\theta_{\mathrm{r},k}^{L_{\mathrm{r}}}\cos\phi_{\mathrm{r},k}^{L_{\mathrm{r}}} \right)^T\right\}, \\
		\mathbf{\Delta}_{k,y_{\mathrm{r}}} \triangleq& \operatorname{Diag}\left\{ \frac{j2\pi}{\lambda} \left(
		\cos\theta_{\mathrm{r},k}^{1},\cdots,\cos\theta_{\mathrm{r},k}^{L_{\mathrm{r}}} \right)^T\right\}.
	\end{aligned}
\end{equation}
Then, function $\mathbf{J}_k^+\left(\mathbf{r}_k\right)$ is  linearized w.r.t. $\mathbf{r}_k$ in $\mathbf{r}_k^{\ell,\hat{\ell}_{\mathrm{r}}}$ by first-order Taylor expression, which leads to
\begin{equation}
	\begin{aligned}
		&\hat{\mathbf{J}}_k^+\left(\mathbf{r}_k\right) \triangleq\mathbf{J}_k^+\left(\mathbf{r}_k^{\ell,\hat{\ell}_{\mathrm{r}}}\right)+\mathbf{\Delta}_{k,X_{\mathrm{r}}}\operatorname{Diag}\left\{\mathbf{x}_{\mathrm{r},k}-\mathbf{x}_{\mathrm{r},k}^{\ell,\hat{\ell}_{\mathrm{r}}}\right\} \\
		&+\operatorname{Diag}\left\{\mathbf{x}_{\mathrm{r},k}-\mathbf{x}_{\mathrm{r},k}^{\ell,\hat{\ell}_{\mathrm{r}}}\right\}\mathbf{\Delta}_{k,X_{\mathrm{r}}}^H +\mathbf{\Delta}_{k,Y_{\mathrm{r}}}\operatorname{Diag}\left\{\mathbf{y}_{\mathrm{r},k}-\mathbf{y}_{\mathrm{r},k}^{\ell,\hat{\ell}_{\mathrm{r}}}\right\} \\
		&+\operatorname{Diag}\left\{\mathbf{y}_{\mathrm{r},k}-\mathbf{y}_{\mathrm{r},k}^{\ell,\hat{\ell}_{\mathrm{r}}}\right\}\mathbf{\Delta}_{k,Y_{\mathrm{r}}}^H.
	\end{aligned}
\end{equation}
Therefore, the concave surrogate of $\bar{R}_{\mathrm{r},k}^+\left(\mathbf{r}_k\right)$ w.r.t. $\mathbf{r}_k$ can be constructed as quadratic-form function $\hat{R}_{J,k}^+\left(\hat{\mathbf{J}}_k^+\left(\mathbf{r}_k\right)\right)$. 
On the other hand, by linearizing $\bar{R}_{\mathrm{r},k}^-\left(\mathbf{r}_k\right)$ w.r.t. $\mathbf{r}_k$ in $\mathbf{r}_k^{\ell,\hat{\ell}_{\mathrm{r}}}$, its first-order Taylor approximation is given by
\begin{equation}
	\begin{aligned}
		\hat{R}_{\mathrm{r},k}^-\left(\mathbf{r}_k\right) = \bar{R}_{\mathrm{r},k}^-\left(\mathbf{r}_k^{\ell,\hat{\ell}_{\mathrm{r}}}\right) + \nabla_{\mathbf{r}_k}^T\bar{R}_{\mathrm{r},k}^-\left(\mathbf{r}_k^{\ell,\hat{\ell}_{\mathrm{r}}}\right) \left( \mathbf{r}_k-\mathbf{r}_k^{\ell,\hat{\ell}_{\mathrm{r}}} \right),
	\end{aligned}
\end{equation}
where $\mathbf{E}_k^-=\left( \mathbf{I}_M + \mathbf{F}_k\left( \mathbf{r}_k \right)^H  \mathbf{\Gamma}_k^- \mathbf{F}_k \left( \mathbf{r}_k \right) \right)^{-1}$ and the gradient is given by
	\begin{equation}
		\label{eq: grad Rbar_rk-}
		\begin{aligned}
			&\nabla_{\mathbf{r}_k}\bar{R}_{\mathrm{r},k}^-\left(\mathbf{r}_k\right)=\\&2\operatorname{Re}\left\{
			\begin{gathered}
				\begin{bmatrix} 
					\operatorname{diag}\left\{\mathbf{E}_k^- \mathbf{F}_k\left( \mathbf{r}_k \right)^H  \mathbf{\Gamma}_k^- \mathbf{\Delta}_{k,x_{\mathrm{r}}} \mathbf{F}_k \left( \mathbf{r}_k \right) \right\} \\
					\operatorname{diag}\left\{\mathbf{E}_k^- \mathbf{F}_k\left( \mathbf{r}_k \right)^H  \mathbf{\Gamma}_k^- \mathbf{\Delta}_{k,y_{\mathrm{r}}} \mathbf{F}_k \left( \mathbf{r}_k \right) \right\}
				\end{bmatrix}
			\end{gathered} \right\}.
		\end{aligned}
	\end{equation}
Consequently, the concave surrogate of $\bar{R}_{\mathrm{r},k}\left(\mathbf{r}_k\right)$ w.r.t. $\mathbf{r}_k$ can be formulated as
\begin{equation}
	\begin{aligned}
		\hat{R}_{\mathrm{r},k}^{\ell,\hat{\ell}_{\mathrm{r}}+1}\left(\mathbf{r}_k\right) &\triangleq \hat{R}_{J,k}^+\left(\hat{\mathbf{J}}_k^+\left(\mathbf{r}_k\right)\right) - \hat{R}_{\mathrm{r},k}^-\left(\mathbf{r}_k\right) \\ &- \delta_{\mathrm{r}} \left( \mathbf{r}_k-\mathbf{r}_k^{\ell,\hat{\ell}_{\mathrm{r}}} \right)^T \left( \mathbf{r}_k-\mathbf{r}_k^{\ell,\hat{\ell}_{\mathrm{r}}} \right),
	\end{aligned}
\end{equation}
where $\delta_{\mathrm{r}}$ is a positive real number to ensure that the objective is concave w.r.t. $\mathbf{r}_k$.
Moreover, given $\mathbf{r}_k^{\ell,\hat{\ell}_{\mathrm{r}}}$, we have $\left\|\mathbf{r}_{k,i}-\mathbf{r}_{k,j}\right\|_2\geq\frac{1}{\left\|\mathbf{r}_{k,i}^{\ell,\hat{\ell}_{\mathrm{r}}}-\mathbf{r}_{k,j}^{\ell,\hat{\ell}_{\mathrm{r}}}\right\|_2}\left(\mathbf{r}_{k,i}-\mathbf{r}_{k,j}\right)^T\left(\mathbf{r}_{k,i}^{\ell,\hat{\ell}_{\mathrm{r}}}-\mathbf{r}_{k,j}^{\ell,\hat{\ell}_{\mathrm{r}}}\right)$, $\forall i \neq j$. Thus, the position constraints of the receive MAs in C4$_k$ are replaced with
\begin{equation}
	\label{eq: r constraint surrogate}
	\begin{aligned}
		\frac{1}{\left\|\mathbf{r}_{k,i}^{\ell,\hat{\ell}_{\mathrm{r}}}-\mathbf{r}_{k,j}^{\ell,\hat{\ell}_{\mathrm{r}}}\right\|_2}\left(\mathbf{r}_{k,i}-\mathbf{r}_{k,j}\right)^T\left(\mathbf{r}_{k,i}^{\ell,\hat{\ell}_{\mathrm{r}}}-\mathbf{r}_{k,j}^{\ell,\hat{\ell}_{\mathrm{r}}}\right) \geq D,\quad \forall i \neq j.
	\end{aligned}
\end{equation}

Based on the above discussion, we solve the following convex surrogate problem in iteration $\hat{\ell}_{\mathrm{r}}+1$ of our proposed SCA algorithm as shown in
\begin{equation}
	\label{op: SCA r surrogate}
	\begin{aligned}
		\max_{\mathbf{r}_k} \quad & \hat{R}_{\mathrm{r},k}^{\ell,\hat{\ell}_{\mathrm{r}}+1}\left(\mathbf{r}_k\right) \\
		\text { s.t. } \quad
		&\text{C3}_k,\eqref{eq: r constraint surrogate}. \\
	\end{aligned}
\end{equation}
To further reduce computational complexity, we consider a problem similar with \eqref{op: SCA r surrogate} but without the constraints, i.e., $\max_{\mathbf{r}_k} \, \hat{R}_{\mathrm{r},k}^{\ell,\hat{\ell}_{\mathrm{r}}+1}\left(\mathbf{r}_k\right)$. The above problem is optimally solved when the first-order optimality condition is satisfied.
Note that the first-order optimality condition is in fact a system of linear equations as in \eqref{eq: grad Rhat_rk} at the bottom of the next page and we have solution
\begin{equation}
	\label{eq: temp optimal r}
	\begin{aligned}
		\tilde{\mathbf{r}}_k^{\ell,\hat{\ell}_{\mathrm{r}}+1} = \mathbf{r}_k^{\ell,\hat{\ell}_{\mathrm{r}}} + \left(\mathbf{A}_k^{\ell,\hat{\ell}_{\mathrm{r}}}\right)^{-1} \mathbf{b}_k^{\ell,\hat{\ell}_{\mathrm{r}}},
	\end{aligned}
\end{equation}
where $
\mathbf{b}_k^{\ell,\hat{\ell}_{\mathrm{r}}} = 2\operatorname{Re}\left\{\left[ \operatorname{diag}\left\{ \mathbf{E}_k^+ \mathbf{\Delta}_{k,X_{\mathrm{r}}} \right\}; \operatorname{diag}\left\{ \mathbf{E}_k^+ \mathbf{\Delta}_{k,Y_{\mathrm{r}}} \right\} \right]\right\} - \nabla_{\mathbf{r}_k}\bar{R}_{\mathrm{r},k}^-\left(\mathbf{r}_k\right)$ and $\mathbf{A}_k^{\ell,\hat{\ell}_{\mathrm{r}}}$ is a non-singular matrix written as \eqref{eq: A_k^ellellhat} at the bottom of the next page.
\begin{figure*}[!b]
	\hrule
	\begin{equation}
		\label{eq: grad Rhat_rk}
		\begin{aligned}
			\nabla_{\mathbf{r}_k}\hat{R}_{\mathrm{r},k}^{\ell,\hat{\ell}_{\mathrm{r}}+1}\left(\mathbf{r}_k\right) =& \operatorname{Re}\left\{
			\begin{gathered}
				\begin{bmatrix} 
					\operatorname{diag}\left\{ -4\mathbf{E}_k^+ \left( \mathbf{J}_k^+\left(\mathbf{r}_k\right)-\mathbf{J}_k^+\left(\mathbf{r}_k^{\ell,\hat{\ell}_{\mathrm{r}}}\right) \right) \mathbf{E}_k^+ \mathbf{\Delta}_{k,X_{\mathrm{r}}} +2\mathbf{E}_k^+ \mathbf{\Delta}_{k,X_{\mathrm{r}}} \right\} \\
					\operatorname{diag}\left\{ 	-4\mathbf{E}_k^+ \left( \mathbf{J}_k^+\left(\mathbf{r}_k\right)-\mathbf{J}_k^+\left(\mathbf{r}_k^{\ell,\hat{\ell}_{\mathrm{r}}}\right) \right) \mathbf{E}_k^+ \mathbf{\Delta}_{k,Y_{\mathrm{r}}} +2\mathbf{E}_k^+ \mathbf{\Delta}_{k,Y_{\mathrm{r}}} \right\} 
				\end{bmatrix}
			\end{gathered}
			\right\} \\-&2\delta_r\left( 	\mathbf{r}_k-\mathbf{r}_k^{\ell,\hat{\ell}_{\mathrm{r}}} \right) - \nabla_{\mathbf{r}_k}\bar{R}_{\mathrm{r},k}^-\left(\mathbf{r}_k\right)=\mathbf{0}.
		\end{aligned}
	\end{equation}
	\begin{equation}
		\label{eq: A_k^ellellhat}
		\begin{aligned}
			\mathbf{A}_k^{\ell,\hat{\ell}_{\mathrm{r}}} =& 	4\operatorname{Re}\left\{ \begin{gathered}
				\begin{bmatrix} 
					\left(\mathbf{E}_k^+ 	\mathbf{\Delta}_{k,X_{\mathrm{r}}}\right) \circ 	\left(\mathbf{E}_k^+ \mathbf{\Delta}_{k,X_{\mathrm{r}}}\right)^T, \left(\mathbf{E}_k^+ \mathbf{\Delta}_{k,Y_{\mathrm{r}}}\right) \circ \left(\mathbf{E}_k^+ \mathbf{\Delta}_{k,X_{\mathrm{r}}}\right)^T \\ 
					\left(\mathbf{E}_k^+ 	\mathbf{\Delta}_{k,X_{\mathrm{r}}}\right) \circ 	\left(\mathbf{E}_k^+ \mathbf{\Delta}_{k,Y_{\mathrm{r}}}\right)^T, \left(\mathbf{E}_k^+ \mathbf{\Delta}_{k,Y_{\mathrm{r}}}\right) \circ \left(\mathbf{E}_k^+ \mathbf{\Delta}_{k,Y_{\mathrm{r}}}\right)^T
				\end{bmatrix}
			\end{gathered} \right.\\&\left.\qquad+ 
			\begin{gathered}
				\begin{bmatrix}  
					\mathbf{E}_k^+ \circ 	\left(\mathbf{\Delta}_{k,X_{\mathrm{r}}}^H\mathbf{E}_k^+ \mathbf{\Delta}_{k,X_{\mathrm{r}}}\right)^T, \mathbf{E}_k^+ \circ \left(\mathbf{\Delta}_{k,Y_{\mathrm{r}}}^H\mathbf{E}_k^+ \mathbf{\Delta}_{k,X_{\mathrm{r}}}\right)^T \\ 
					\mathbf{E}_k^+ \circ 	\left(\mathbf{\Delta}_{k,X_{\mathrm{r}}}^H\mathbf{E}_k^+ \mathbf{\Delta}_{k,Y_{\mathrm{r}}}\right)^T, \mathbf{E}_k^+ \circ \left(\mathbf{\Delta}_{k,Y_{\mathrm{r}}}^H\mathbf{E}_k^+ \mathbf{\Delta}_{k,Y_{\mathrm{r}}}\right)^T
				\end{bmatrix}
			\end{gathered} 
			\right\}+2\delta_{\mathrm{r}}\mathbf{I}_{2M}.\\
		\end{aligned}
	\end{equation} 
\end{figure*}
After that, if $\tilde{\mathbf{r}}_k^{\ell,\hat{\ell}_{\mathrm{r}}+1}$ satisfies constraints C3$_k$ and \eqref{eq: r constraint surrogate}, it is the optimal solution to problem \eqref{op: SCA r surrogate}. Otherwise, problem \eqref{op: SCA r surrogate} can be efficiently solved by traditional convex optimization techniques, such as CVX. Then, the receive APV is updated
according to
\begin{equation}
	\label{eq: update r}
	\begin{aligned}
		\mathbf{r}_k^{\ell,\hat{\ell}_{\mathrm{r}}+1}=\mathbf{r}_k^{\ell,\hat{\ell}_{\mathrm{r}}}+\tau_{\mathrm{r}}^{\ell,\hat{\ell}_{\mathrm{r}}+1} \left(\bar{\mathbf{r}}_k^{\ell,\hat{\ell}_{\mathrm{r}}+1}-\mathbf{r}_k^{\ell,\hat{\ell}_{\mathrm{r}}}\right),
	\end{aligned}
\end{equation}
where $\bar{\mathbf{r}}_k^{\ell,\hat{\ell}_{\mathrm{r}}+1}$ and $\tau_{\mathrm{r}}^{\ell,\hat{\ell}_{\mathrm{r}}+1}$ are the solution to problem \eqref{op: SCA r surrogate} and the step size selected by backtracking line search, respectively. In each iteration, we start with a step size, $\tau_{\mathrm{r}}^{\ell,\hat{\ell}_{\mathrm{r}}+1}=\tau^0$, and repeatedly reduce it to $\kappa\tau_{\mathrm{r}}^{\ell,\hat{\ell}_{\mathrm{r}}+1}$, until the following Armijo–Goldstein condition is satisfied:
\begin{equation}
\label{eq: Armijo–Goldstein condition r}
\begin{aligned}
	 I_{\mathrm{r}}^{\ell}\left(\mathbf{r}_k^{\ell,\hat{\ell}_{\mathrm{r}}+1},\mathbf{r}_k^{\ell,\hat{\ell}_{\mathrm{r}}}\right) \geq \xi \tau_{\mathrm{r}}^{\ell,\hat{\ell}_{\mathrm{r}}+1} \left\|	\bar{\mathbf{r}}_k^{\ell,\hat{\ell}_{\mathrm{r}}+1}-\mathbf{r}_k^{\ell,\hat{\ell}_{\mathrm{r}}} \right\|^2,
\end{aligned}
\end{equation}
where $I_{\mathrm{r}}^{\ell}\left(\mathbf{r}_k^{\ell,\hat{\ell}_{\mathrm{r}}+1},\mathbf{r}_k^{\ell,\hat{\ell}_{\mathrm{r}}}\right)\triangleq \bar{R}_{\mathrm{r},k}\left(\mathbf{r}_k^{\ell,\hat{\ell}_{\mathrm{r}}+1}\right) - \bar{R}_{\mathrm{r},k}\left(\mathbf{r}_k^{\ell,\hat{\ell}_{\mathrm{r}}}\right)$ is the increment of the objective of problem \eqref{op: SR r}. The SCA algorithm runs until $\hat{\ell}_{\mathrm{r}} = \hat{L}_{\mathrm{r}}$ and the increment is less than $\epsilon_1$, i.e.,
\begin{equation}
	\label{eq: termin condition r}
	\begin{aligned}
		I_{\mathrm{r}}^{\ell}\left(\mathbf{r}_k^{\ell,\hat{\ell}_{\mathrm{r}}+1},\mathbf{r}_k^{\ell,\hat{\ell}_{\mathrm{r}}}\right) \leq \epsilon_1.
	\end{aligned}
\end{equation}

The overall SCA algorithm is summarized in Algorithm \ref{Al: SCA r}. In particular, $\mathbf{r}_k^{\ell,0}$ is initialized as the output of the AO algorithm in the last iteration (i.e., $\mathbf{r}_k^{\ell-1}$) in Step 1 and then optimized by the SCA algorithm in Steps 2-15. In Steps 3-8, problem \eqref{op: SCA r surrogate} is constructed and solved to obtain solution $\bar{\mathbf{r}}_k^{\ell,\hat{\ell}_{\mathrm{r}}+1}$. In Steps 9-14, the transmit APV $\mathbf{r}_k^{\ell,\hat{\ell}_{\mathrm{r}}+1}$ is updated with proper step size $\tau_{\mathrm{r}}^{\ell,\hat{\ell}_{\mathrm{r}}+1}$ chosen by backtracking line search. Since the convergence analysis of Algorithm \ref{Al: SCA r} is similar to Algorithm \ref{Al: SCA t}, it is omitted here for brevity. In addition, the computational complexity of Algorithm \ref{Al: SCA r} is mainly determined by solving problem \eqref{op: SCA r surrogate} in each iteration, which entails $\mathcal{O}\left(M^{3.5}\log(1/\epsilon)\right)$ when the interior-point method is employed with accuracy $\epsilon$. As a result, the overall computational complexity of Algorithm \ref{Al: SCA r} is $\mathcal{O}\left(\hat{L}_{\mathrm{r}}M^{3.5}\log(1/\epsilon)\right)$.
\begin{algorithm}[t]
	\caption{Algorithm for Solving Problem \eqref{op: SR r}}
	\label{Al: SCA r}
	\begin{spacing}{0.9}
		\KwIn{$M$, $\lambda$, $D$, $\mathcal{C}_{\mathrm{r},k}$; $\delta_{\mathrm{r}}$, $\epsilon_1$, $\hat{L}_{\mathrm{r}}$, $\tau^0$, $\kappa$, $\xi$}
		\KwOut{$\mathbf{r}_k^{\ell}$}
		
		Initialize $\hat{\ell}_{\mathrm{r}}=0$ and $\mathbf{r}_k^{\ell,0}=\mathbf{r}_k^{\ell-1}$
			
		\Repeat{{\eqref{eq: termin condition r} is satisfied and $\hat{\ell}_{\mathrm{r}} = \hat{L}_{\mathrm{r}}$}}{
			
			Calculate $\tilde{\mathbf{r}}_k^{\ell,\hat{\ell}_{\mathrm{r}}+1}$ by \eqref{eq: temp optimal r}\
			
			\eIf{$\tilde{\mathbf{r}}_k^{\ell,\hat{\ell}_{\mathrm{r}}+1}$ satisfyies constraints $\text{C4}_k$ and \eqref{eq: r constraint surrogate}}
			{Set $\bar{\mathbf{r}}_k^{\ell,\hat{\ell}_{\mathrm{r}}+1}=\tilde{\mathbf{r}}_k^{\ell,\hat{\ell}_{\mathrm{r}}+1}$}
			{Calculate $\bar{\mathbf{r}}_k^{\ell,\hat{\ell}_{\mathrm{r}}+1}$ by solving problem \eqref{op: SCA r surrogate}}
			
			Initialize $\tau_{\mathrm{r}}^{\ell,\hat{\ell}_{\mathrm{r}}+1}=\tau^0/\kappa$\
			
			\Repeat{{\eqref{eq: Armijo–Goldstein condition r} is satisfied}}{
			Set $\tau_{\mathrm{r}}^{\ell,\hat{\ell}_{\mathrm{r}}+1}=\kappa\tau_{\mathrm{r}}^{\ell,\hat{\ell}_{\mathrm{r}}+1}$\
			
			Update $\mathbf{r}_k^{\ell,\hat{\ell}_{\mathrm{r}}+1}$ by \eqref{eq: update r}\
			}
			
			Set $\hat{\ell}_{\mathrm{r}} = \hat{\ell}_{\mathrm{r}}+1$\
			}
			
			\textbf{return} $\mathbf{r}_k^{\ell}=\mathbf{r}_k^{\ell,\hat{L}_{\mathrm{r}}}$\
	\end{spacing}
\end{algorithm}

\subsection{Complexity and Convergence Analysis}
The overall AO algorithm is summarized in Algorithm \ref{Al: AO}. In Step 1, we initialize $\mathbf{t}^0$, $\left\{\mathbf{r}_k^0\right\}$ and $\mathbf{P}^0$, satisfying all the constraints and then they are alternatively optimized in Steps 2-12. In each iteration, transmit variables $\mathbf{t}^{\ell}$ and $\mathbf{P}^{\ell}$ are first calculated by Algorithm \ref{Al: SCA t} after $\left\{\hat{R}_{\mathrm{t},k}^{\ell+1}\left(\mathbf{t},\mathbf{P}\right)\right\}$ is constructed through the DE iterative process and \eqref{eq: DE minorizing function t} in Steps 3-8. Subsequently, receive APVs $\left\{\mathbf{r}_k^{\ell}\right\}$ are calculated by Algorithm \ref{Al: SCA r} after constructing $\left\{\bar{R}_{\mathrm{r},k}\left(\mathbf{r}_k\right)\right\}$ through the DE iterative process and \eqref{eq: DE Rrk}.

The convergence of Algorithm \ref{Al: AO} is analyzed as follows. Note that the objectives of problems \eqref{op: EE t P} and \eqref{op: SR r} are the tight minorizing function of the original objective and the tight approximation of the corresponding average achievable rate, respectively. From the convergence analysis of Algorithms \ref{Al: SCA t} and \ref{Al: SCA r}, it is apparent that the proposed AO algorithm generates a non-decreasing sequence $\left\{\mathbf{t}^0, \left\{\mathbf{r}_k^0\right\}, \mathbf{P}^0\right\}, \left\{\mathbf{t}^1, \left\{\mathbf{r}_k^1\right\}, \mathbf{P}^1\right\}, \cdots$. Besides, the objective of problem \eqref{op: original} is upper-bounded by a finite value since the feasible region is compact. Thus, the convergence of Algorithm \ref{Al: AO} is guaranteed. 
On the other hand, the computational complexity of a DE iterative process is $\mathcal{O}\left(L_{\mathrm{de}}M^{3}+L_{\mathrm{de}}\min\left(N^{3},L_{\mathrm{t}}^{3}\right)\right)$, where $L_{\mathrm{de}}$ is the number of iterations.
Consequently, the computational complexity of Algorithm \ref{Al: AO} is $\mathcal{O}\left( L_{\mathrm{ao}} \left( L_{\mathrm{de}}M^{3}+L_{\mathrm{de}}\min\left(N^{3},L_{\mathrm{t}}^{3}\right)+\hat{L}_{\mathrm{t}}N^{3.5}\log(1/\epsilon)+\right.\right.$$\\ \left.\left.\hat{L}_{\mathrm{r}}KM^{3.5}\log(1/\epsilon) \right) \right)$.
\begin{algorithm}[t]
	\caption{AO Algorithm for Solving Problem \eqref{op: original}}
	\label{Al: AO}
	\begin{spacing}{0.9}
		\KwIn{$N$, $M$, $K$, $\sigma^2$, $\lambda$, $D$, $\mathcal{C}_{\mathrm{t}}$, $\{\mathcal{C}_{\mathrm{r},k}\}$, $\omega$, $P_{\mathrm{max}}$, $P_{\mathrm{c}}$, $P_{\mathrm{s}}$; $\epsilon_1$, $L_{\mathrm{ao}}$, $L_{\mathrm{de}}$, $\hat{L}_{\mathrm{t}}$, $\hat{L}_{\mathrm{r}}$; S-CSI}
		\KwOut{$\mathbf{t}^{L_{\mathrm{ao}}}\text{, }\left\{\mathbf{r}_k^{L_{\mathrm{ao}}}\right\}\text{ and }\mathbf{P}^{L_{\mathrm{ao}}}$}
		
		Initialize $\mathbf{t}^0$, $\left\{\mathbf{r}_k^0\right\}$ and $\mathbf{P}^0$, satisfying all the constraints
		
		\For{$\ell = 1 \text{ to } L_{\mathrm{ao}}$}{
			\For{$\ell = 1 \text{ to } L_{\mathrm{de}}$}{
			Calculate $\left\{\widetilde{\mathbf{\Phi}}_k^+\right\}$ and $\left\{\mathbf{\Phi}_k^+\right\}$ ($\left\{\widetilde{\mathbf{\Phi}}_k^-\right\}$ and $\left\{\mathbf{\Phi}_k^-\right\}$) by \eqref{eq: DE proces (a)} and \eqref{eq: DE proces (b)} (with $\mathbf{Q}\leftarrow\mathbf{Q}_{\backslash k}$)\
			
			Calculate $\left\{\widetilde{\mathbf{\Theta}}_k^+\right\}$ and $\left\{\mathbf{\Theta}_k^+\right\}$ ($\left\{\widetilde{\mathbf{\Theta}}_k^-\right\}$ and $\left\{\mathbf{\Theta}_k^-\right\}$) by \eqref{eq: DE proces (c)} and \eqref{eq: DE proces (d)} (with $\mathbf{Q}\leftarrow\mathbf{Q}_{\backslash k}$)\
		}
			Construct $\left\{\hat{R}_{\mathrm{t},k}^{\ell+1}\left(\mathbf{t},\mathbf{P}\right)\right\}$ by \eqref{eq: DE minorizing function t}\
			
			Calculate $\mathbf{t}^{\ell}$ and $\mathbf{P}^{\ell}$ by Algorithm \ref{Al: SCA t}\
						
			Execute Steps 3-6\
			
			Construct $\left\{\bar{R}_{\mathrm{r},k}\left(\mathbf{r}_k\right)\right\}$ by \eqref{eq: DE Rrk}\
			
			Calculate $\left\{\mathbf{r}_k^{\ell}\right\}$ by Algorithm \ref{Al: SCA r}\
			
		}
			\textbf{return} $\mathbf{t}^{L_{\mathrm{ao}}}$, $\left\{\mathbf{r}_k^{L_{\mathrm{ao}}}\right\}$ and $\mathbf{P}^{L_{\mathrm{ao}}}$\
	\end{spacing}
\end{algorithm}

\begin{Remark}
	So far, the AO algorithm is confined to the multi-user scenario. For the single-user system, the objective of problem \eqref{op: EE t P} is replaced by the approximation of the average achievable rate, $\frac{ \bar{R}_{\mathrm{t},1}^+\left(\mathbf{t},\mathbf{P}_1\right) } {P_\mathrm{tot}}$. Meanwhile, an iterative algorithm for solving problem \eqref{op: EE P} with the replaced objective can be developed similar to that for the multi-user scenario, since $\frac{ \bar{R}_{\mathrm{t},1}^+\left(\mathbf{t},\mathbf{P}_1\right) } {P_\mathrm{tot}}$ is concave-convex fractional w.r.t. $\mathbf{P}_1$ for fixed $\mathbf{t}$.
\end{Remark}

\section{Numerical Results}
In this section, we numerically evaluate the performance of the proposed multi-user MA-enhanced MIMO system. We first introduce the simulation setup and benchmark schemes. Subsequently, we present numerical results to verify the efficacy of the proposed AO algorithm.

\subsection{Simulation Setup and Benchmark Schemes}
In our simulations, the UTs are randomly distributed around the BS with their distances w.r.t. the BS, $d_k$, following a uniform distribution between $20$ and $100$ meters (m). For UT $k$, the channel gain is set to $g_k = c_0 d_k^{-\alpha_0}$, where $c_0$ denotes the expected value of the path loss at the reference distance of $1$ m, and $\alpha_0$ represents the path loss exponent.
Moreover, we assume that there are $L_{\mathrm{t}}=L_{\mathrm{r}}=L$ paths for each UT and the entries of $\overline{\mathbf{\Sigma}}_{k}$ are set to $0$ except $\left[\overline{\mathbf{\Sigma}}_{k}\right]_{1,1} = \sqrt{ \frac{g_k K_\mathrm{r}}{\left(K_\mathrm{r}+1\right)} }$ as well as the entries of $\mathbf{M}_{k}$ are set to $0$ except $\left[\mathbf{M}_{k}\right]_{l,l} = \sqrt{ \frac{g_k}{\left(L-1\right)\left(K_\mathrm{r}+1\right)} }$, $2 \leq l \leq L$, where $K_\mathrm{r}$ is the Rician factor. Moreover, the $L$ pairs of elevation/azimuth AoDs and AoAs for UT $k$ are i.i.d. random variables following distributions $f\left(\theta_{\mathrm{t},k}^l,\phi_{\mathrm{t},k}^l\right)=\frac{1}{2\pi} \sin \phi_{\mathrm{t},k}^l$, $\theta_{\mathrm{t},k}^l\in\left[0,\pi\right]$, $\phi_{\mathrm{t},k}^l\in\left[0,\pi\right]$ and $f\left(\theta_{\mathrm{r},k}^l,\phi_{\mathrm{r},k}^l\right)=\frac{1}{2\pi} \sin \phi_{\mathrm{r},k}^l$, $\theta_{\mathrm{r},k}^l\in\left[0,\pi\right]$, $\phi_{\mathrm{r},k}^l\in\left[0,\pi\right]$, $1\leq l \leq L$, respectively. 
\begin{table}[htb]
	\begin{center}
		\caption{Simulation Parameters.}
		\label{Tb: simulation parameter}
		\begin{tabular}{|c|c|c|c|}
			\hline Parameter & Value & Parameter & \,\, Value \,\, \\
			\hline $N$ & $16$ & $\omega$ & $5$ \\
			\hline $M$, $s_k$ & $4$ & $P_{\mathrm{max}}$ & $30\mathrm{~dBm}$ \\
			\hline $K$ & $4$ & $P_{\mathrm{c}}$ & $30\mathrm{~dBm}$ \\
			\hline $\lambda$ & $10~\mathrm{cm}$ & $P_{\mathrm{s}}$ & $40\mathrm{~dBm}$ \\
			\hline $D$ & $0.5\lambda$ & $L_{\mathrm{ao}}$ & $50$ \\
			\hline $X_{\mathrm{t}}$ & $3.2\lambda$ & $L_{\mathrm{de}}$ & $20$ \\
			
			\hline $X_{\mathrm{r}}$ & $2\lambda$ & $\hat{L}_{\mathrm{t}}$, $\hat{L}_{\mathrm{r}}$ & $20$ \\
			\hline $L$ & $5$ & $\delta_{\mathrm{t}}$, $\delta_{\mathrm{r}}$ & $0.02$ \\
			\hline $K_\mathrm{r}$ & $1$ & $\tau^0$ & $1$ \\
			\hline $\sigma^2$ & $-80\mathrm{~dBm}$ & $\kappa$ & $0.2$ \\
			\hline $c_0$ & $-40\mathrm{~dB}$ & $\xi$ & $0.02$ \\
			\hline $\alpha_0$ & $2.8$ & $\epsilon_1$ & $10^{-3}$ \\
			\hline 
		\end{tabular}
	\end{center}
\end{table}
The size of the transmit and receive regions are set to $X_{\mathrm{t}}\times X_{\mathrm{t}}$ and $X_{\mathrm{r}}\times X_{\mathrm{r}}$, respectively. The default settings for the simulation parameters are provided in Table \ref{Tb: simulation parameter} \cite{Mccf}. We compare the performance of the proposed MA-enhanced system with the following schemes, where the adjacent antennas in the fixed-position UPAs are spaced by $0.5\lambda$:
\begin{itemize}
	\item [$\bullet$]\textbf{CSSCA}: A CSSCA algorithm similar to that in \cite{JBaA} is developed to design the proposed MA-enhanced system.
	\item [$\bullet$]\textbf{TMA}: The MAs are only deployed at the BS and each UT is equipped with $2\times2$ UPA.
	\item [$\bullet$]\textbf{RMA}: The MAs are only deployed at the UTs and the BS is equipped with $4\times4$ UPA.
	\item [$\bullet$]\textbf{Discrete position selection (DPS)}: The transmit and receive regions are quantized into discrete
	$\lfloor \frac{X_{\mathrm{t}}}{D} \rfloor \times \lfloor \frac{X_{\mathrm{t}}}{D} \rfloor$ and $\lfloor \frac{X_{\mathrm{r}}}{D} \rfloor \times \lfloor \frac{X_{\mathrm{r}}}{D} \rfloor$ locations, respectively. In addressing problems \eqref{op: EE t} and \eqref{op: SR r}, we alternatively optimize the position for each MA by the exhaustive search within its current or vertically/horizontally adjacent positions that are unoccupied by other MAs.
	\item [$\bullet$]\textbf{UPA}: The BS and each UT are equipped with $4\times4$ and $2\times2$ UPAs, respectively.
	\item [$\bullet$]\textbf{MA-ICSI}: Instead of the S-CSI, the I-CSI is leveraged to design the proposed MA-enhanced system and evaluate the performance.
	\item [$\bullet$]\textbf{MA-LOS}: The LOS components are leveraged to design the proposed MA-enhanced system, while the S-CSI is still leveraged to evaluate the performance.
\end{itemize}

\subsection{Convergence Behavior}
\begin{figure}
	\centering
	\includegraphics[width=0.45\textwidth,height=0.35\textwidth]{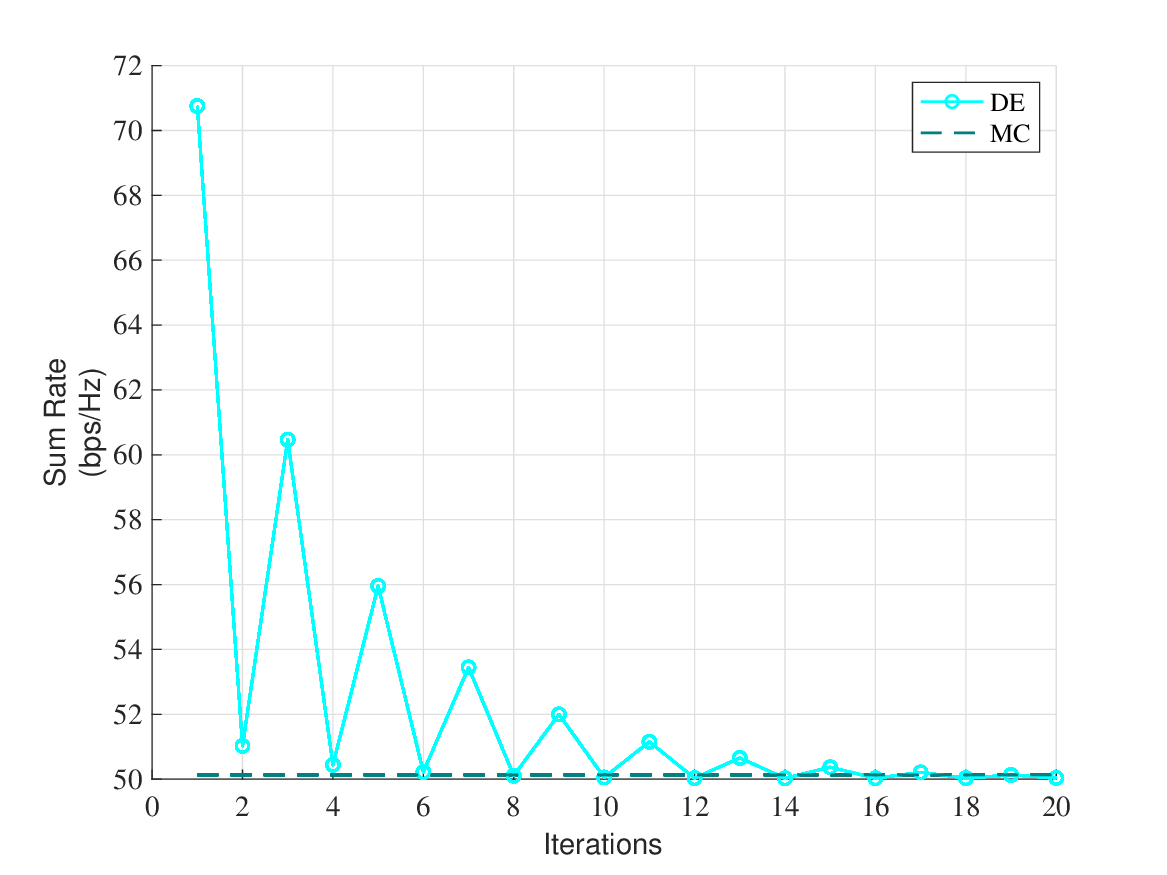}
	\caption{The convergence behavior of the DE technique.}
	\label{fig: DE iteration}
\end{figure}
\begin{figure}
	\centering
	\includegraphics[width=0.45\textwidth,height=0.35\textwidth]{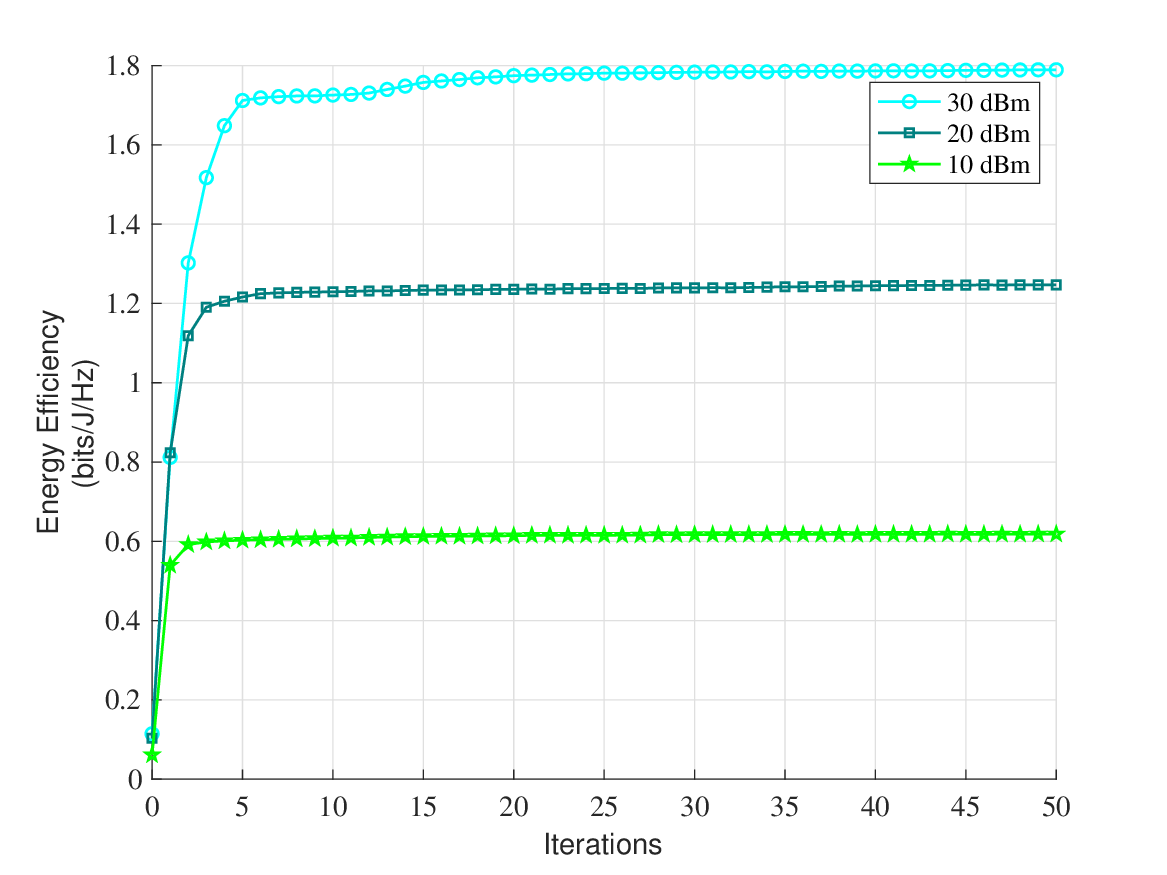}
	\caption{The convergence behavior of the proposed algorithm.}
	\label{fig: iteration}
\end{figure}
As shown in Fig.~\ref{fig: DE iteration}, the average achievable sum rate obtained by the DE technique achieves fast convergence, which approaches the sum rate averaged by the MC simulation after $20$ iterations.
In addition, the convergence behavior of Algorithm \ref{Al: AO} is illustrated in Fig.~\ref{fig: iteration}, where we analyze the EE performance versus the number of iterations under different transmit power thresholds. The results demonstrate that the proposed algorithm exhibits rapid convergence under all considered transmit power thresholds. Notably, the required number of iterations for convergence increases with the transmit power threshold $P_{\mathrm{max}}$. Specifically, when $P_{\mathrm{max}}$ is set to $10~\mathrm{dBm}$, the algorithm attains optimal performance within merely $4$ iterations. In contrast, for a higher transmit power threshold $P_{\mathrm{max}}=30~\mathrm{dBm}$, approximately $25$ iterations are needed to achieve convergence.

\subsection{Impact of the Maximum Transmit Power}
\begin{figure}
	\centering
	\includegraphics[width=0.5\textwidth,height=0.4\textwidth]{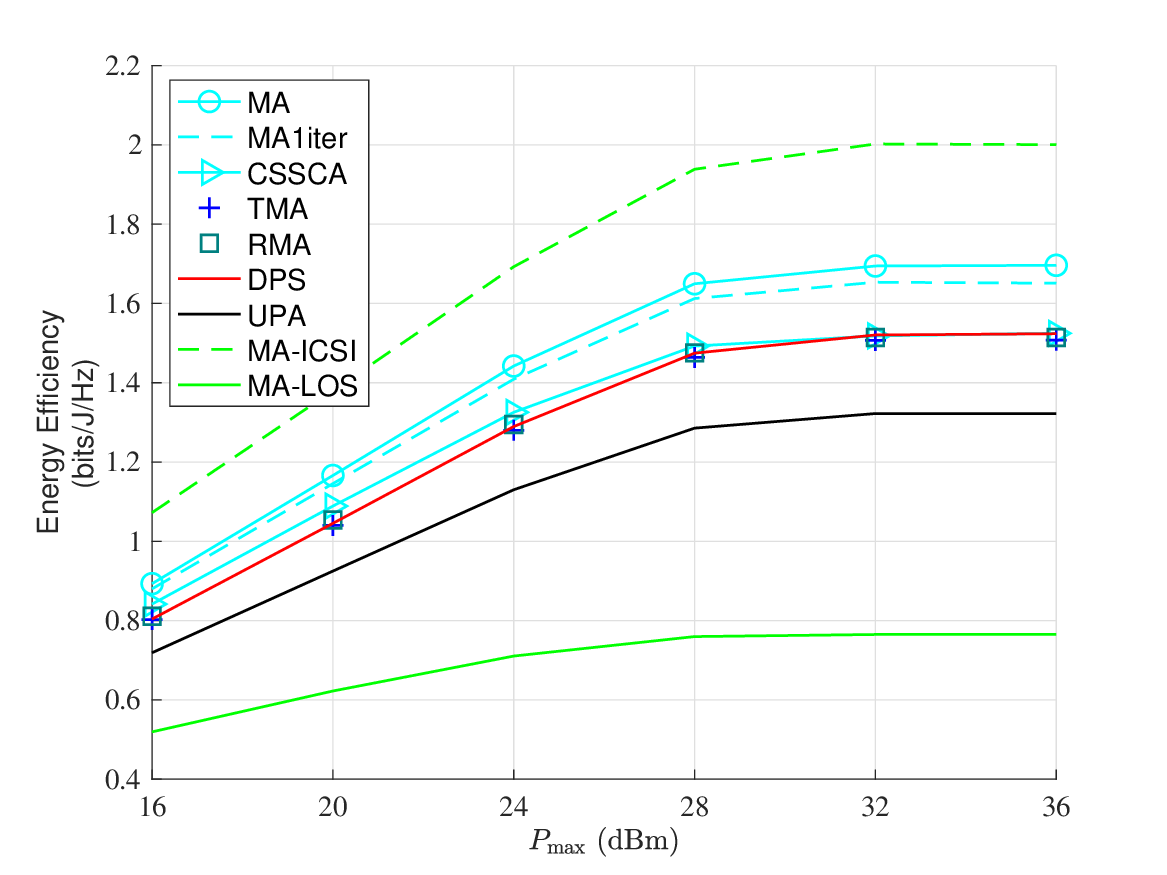}
	\caption{EE versus the maximum transmit power.}
	\label{fig: Simulation P}
\end{figure}
From Fig. \ref{fig: Simulation P}, it can be seen that the system EE increases with the maximum transmit power and finally reaches saturation when $P_{\mathrm{max}}$ surpasses $32 \mathrm{~dBm}$ for the considered schemes. This phenomenon occurs because
any transmit power that surpasses the optimal threshold will not yield any further enhancement in EE performance. Additionally, the results highlight that the MA-enhanced (except MA-LOS) schemes exhibit superior performance relative to the UPA scheme, owing to the distinct amounts of spatial degrees of freedom (DoFs) exploited by each. The ideal MA-ICSI scheme achieves the optimal EE performance since the optimization variables are specially designed based on the I-CSI in each coherence time interval. In contrast, the MA-LOS scheme achieves the worst EE performance since only the LOS components are exploited for system design, while the NLOS components are not leveraged to enhance the desired signals or mitigate the interference. Notably, our proposed MA system demonstrates the most outstanding EE performance amongst all the schemes utilizing the S-CSI, irrespective of the power level. We also consider the MA1iter scheme, which is similar to the proposed MA scheme, except that the maximum number of iterations of the SCA algorithms is set to $1$, i.e., $\hat{L}_{\mathrm{t}}=\hat{L}_{\mathrm{r}}=1$. While the SCA algorithms are executed for only one iteration, it achieves a superior EE performance, which is extremely close to the proposed MA scheme and surpasses the CSSCA scheme.

\subsection{Impact of the Number of UTs}
\begin{figure}
	\centering
	\includegraphics[width=0.5\textwidth,height=0.4\textwidth]{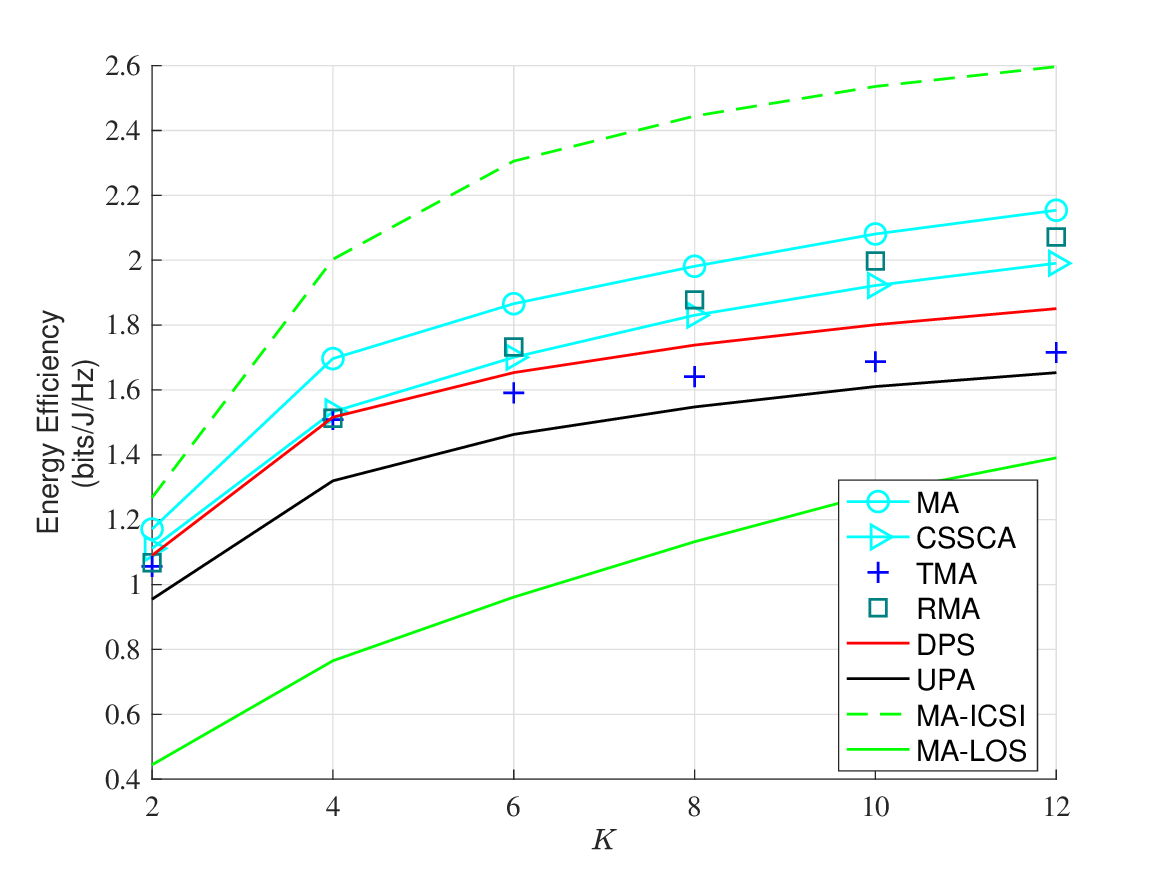}
	\caption{EE versus the number of UTs.}
	\label{fig: Simulation K}
\end{figure}
Fig. \ref{fig: Simulation K} depicts the correlation between the system EE and the number of UTs for the proposed MA system and the benchmark schemes. The system EE improves with $K$ due to the exploitation of the multi-user diversity. For the MA, CSSCA, RMA, DPS, MA-ICSI and MA-LOS schemes, the MAs deployed at the UTs gain access to more spatial DoFs that can be utilized to mitigate the multi-user interference and harness the spatial diversity. As a result, the system EE grows faster than that in the other benchmark schemes.

\subsection{Impact of the Movement Region Size}
\begin{figure}
	\centering
	\includegraphics[width=0.5\textwidth,height=0.4\textwidth]{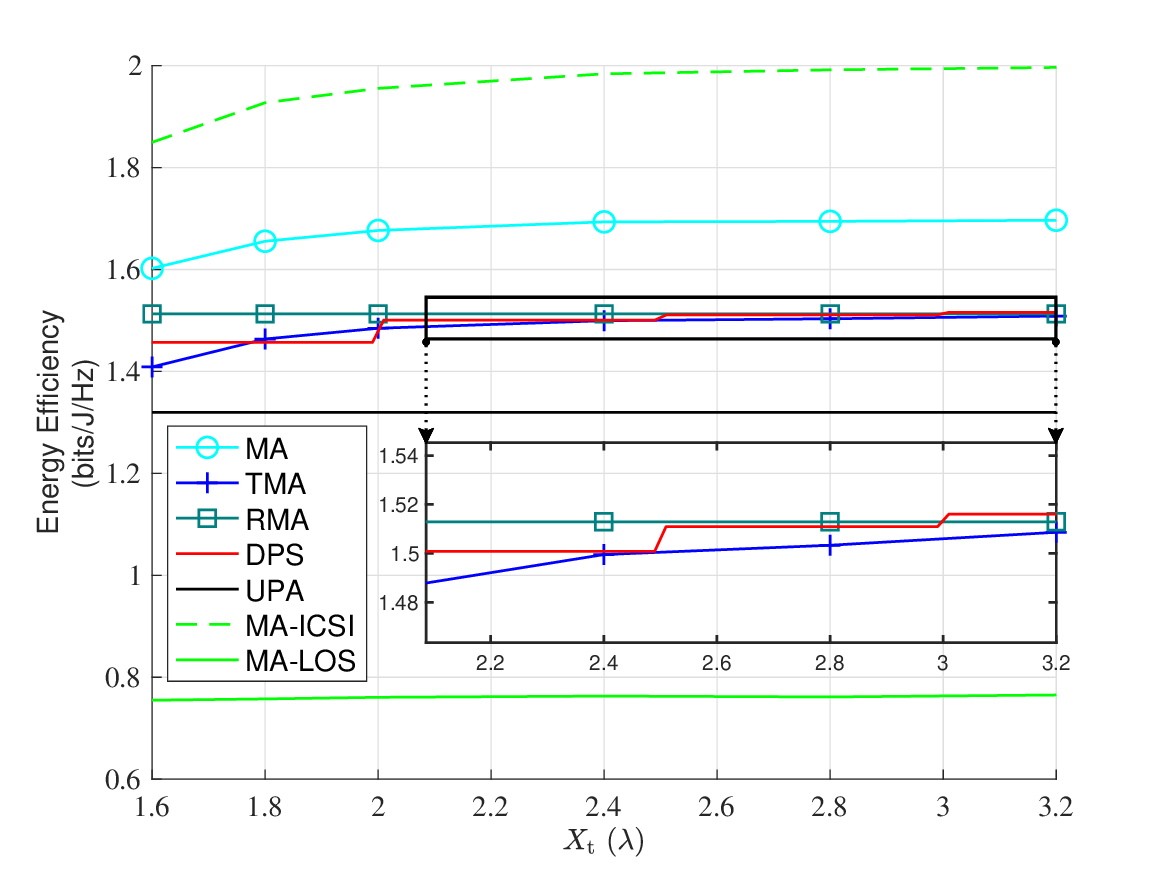}
	\caption{EE versus the size of the transmit region.}
	\label{fig: Simulation T}
\end{figure}
\begin{figure}
	\centering
	\includegraphics[width=0.5\textwidth,height=0.4\textwidth]{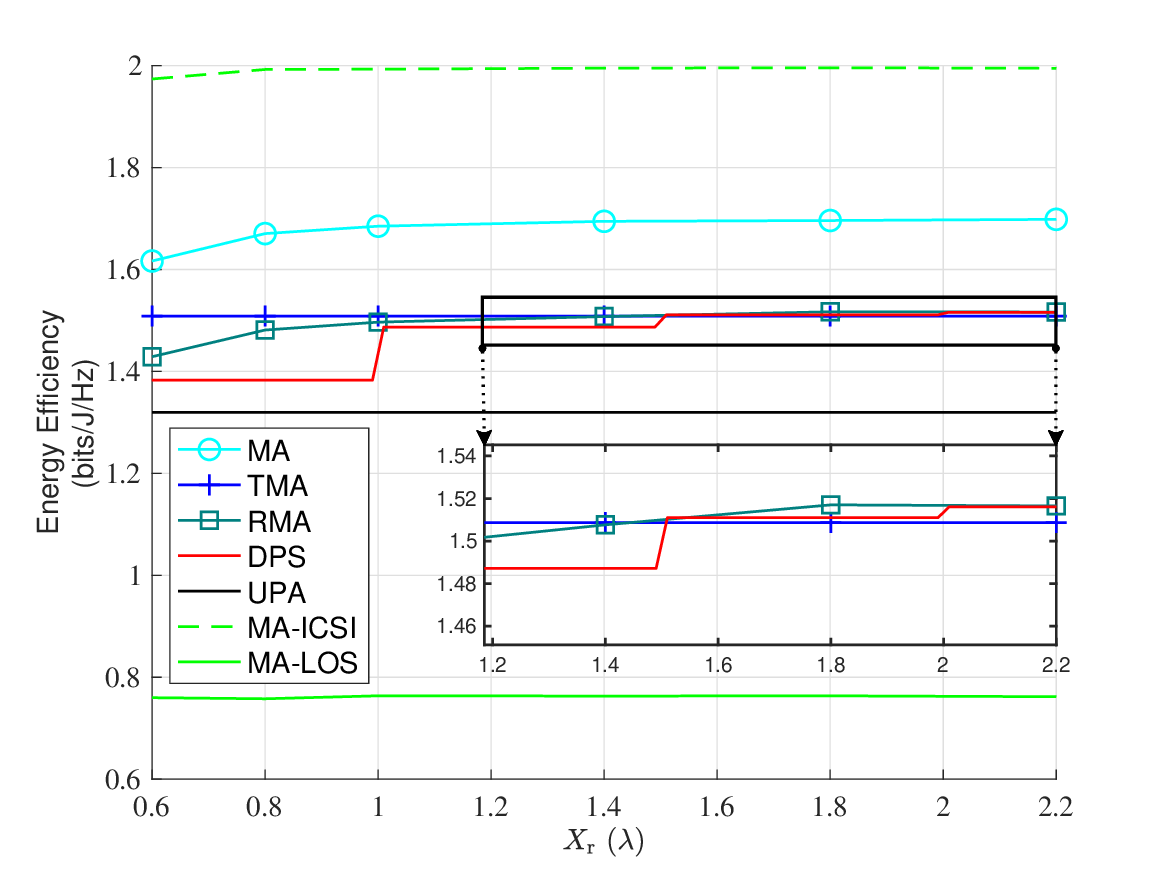}
	\caption{EE versus the size of the receive region.}
	\label{fig: Simulation R}
\end{figure}
Fig. \ref{fig: Simulation T} illustrates the effect of the transmit region size on the system EE performance. As the increase in $X_\mathrm{t}$ enlarges the movement region for the transmit MA design, the system EE of the MA, TMA, DPS and MA-ICSI schemes exhibits a significant upward trend with $X_\mathrm{t}$. Conversely, the MA-LOS scheme achieves a negligible performance gain as the great effect of the NLOS components are neglected. Moreover, all schemes tend to converge when the transmit region size exceeds $3\lambda$, indicating that a nearly optimal system EE can be achieved with a finite transmit region size. With $X_\mathrm{t}=3.2\lambda$, the MA, TMA, RMA, DPS and MA-ICSI schemes achieve $28.5\%$, $14.3\%$, $14.6\%$, $14.9\%$ and $51.3\%$ EE improvements over the UPA scheme, respectively. Fig. \ref{fig: Simulation R} illustrates the system EE versus the size of the receive region. As expected, the system EE of the schemes with receive MAs ascends with $X_\mathrm{r}$ since the receive MAs can be adjusted in a larger region and converges when the receive region size exceeds $2\lambda$. Moreover, with $X_\mathrm{r}=2.2\lambda$, the MA, TMA, RMA, DPS and MA-ICSI schemes achieve $28.7\%$, $14.3\%$, $14.9\%$, $14.9\%$ and $51.1\%$ EE improvements over the UPA scheme, respectively.

\subsection{Impact of the Channel Condition}
\begin{figure*}[h]
	\centering
	\subfigure[EE versus the number of each UT's channel paths.]{\includegraphics[width=0.45\textwidth,height=0.35\textwidth]{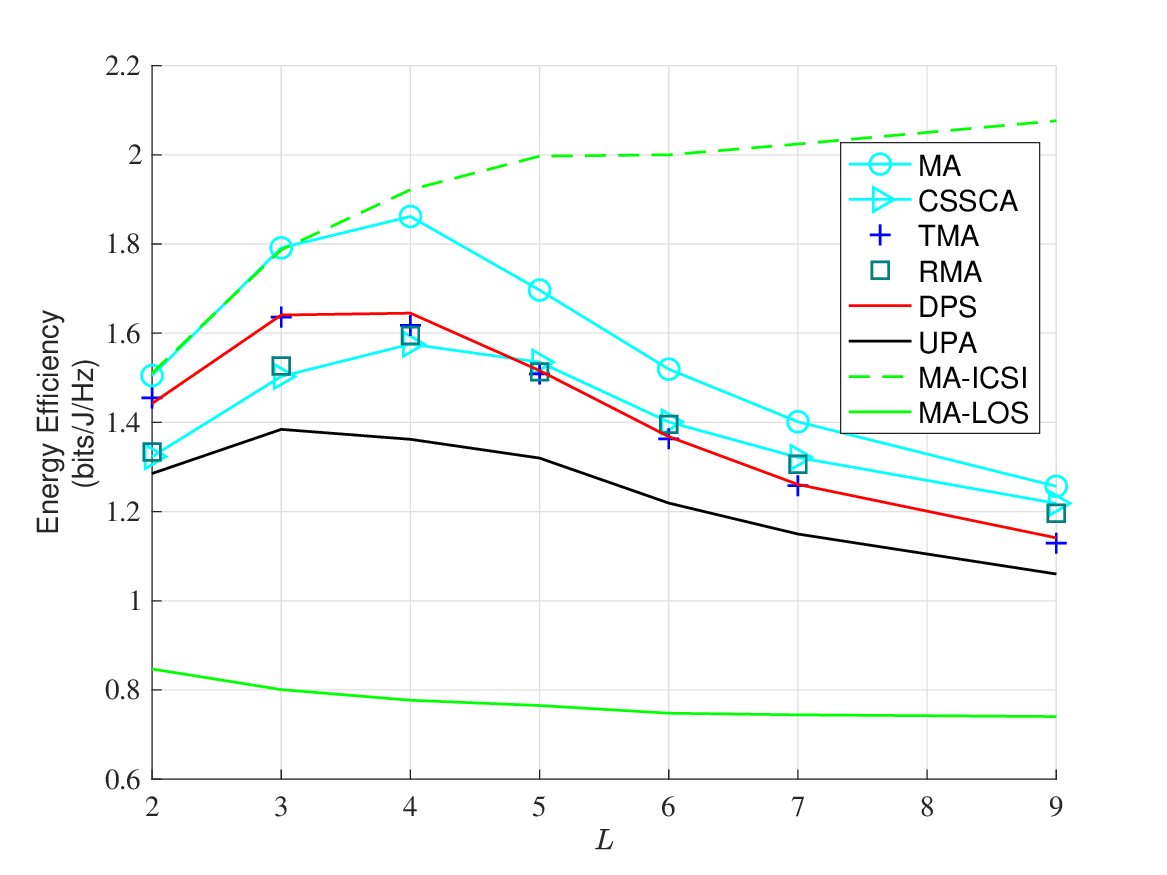}\label{fig: Simulation LtLr}}
	\subfigure[EE versus the Rician factor.]{\includegraphics[width=0.45\textwidth,height=0.35\textwidth]{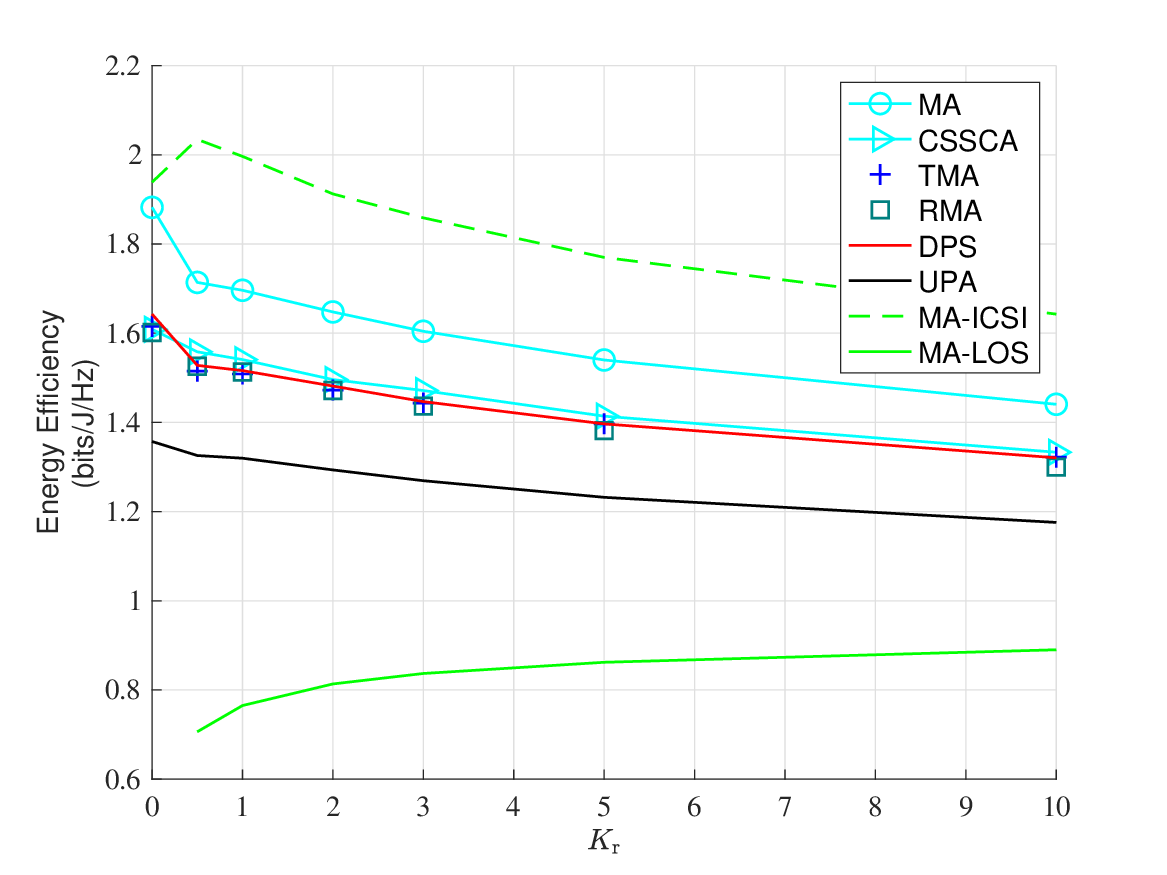}\label{fig: Simulation Kr}}
	\caption{Impact of the channel condition.}
	\label{fig: Simulation Channel Condition}
\end{figure*}
Fig.~\ref{fig: Simulation Channel Condition} depicts the relationship between the system EE and the channel condition. As shown in Fig. \ref{fig: Simulation LtLr}, compared to the TMA and UPA schemes, the system EE of the MA, CSSCA, RMA and DPS schemes exhibits a monotonic upward trend with $L$ when $L$ is less than the number of MAs at each UT, which indicates that the spatial DoFs from various paths are effectively exploited through the adjustment of receive MAs. However, when $L$ becomes large, the system EE of all the above schemes decline with increasing $L$ as channel power is spread over multiple paths. Driven by the same negative factor, this declining trend can be also observed in the MA-LOS scheme throughout the entire range in simulation. Conversely, this negative factor is offset by the adaptability of the MAs in the MA-ICSI scheme, which can be dynamically adjusted according to the realized path gains within each coherence time interval. As a result, more spatial DoFs can be exploited and the system EE exhibits a monotonic increase with the growth of $L$. 
Generally, MAs can provide a more significant performance gain under proper multi-path propagation conditions. Therefore, in Fig. \ref{fig: Simulation Kr}, when $K_\mathrm{r}\geq 0.5$, the system EE for the MA, CSSCA, TMA, RMA, DPS, UPA and MA-ICSI schemes decreases with increasing $K_\mathrm{r}$ as the utilized Rayleigh channel paths are weaken. Moreover, when $K_\mathrm{r}$ increases from $0$ (correspond to four Rayleigh channel paths) to $0.5$ (correspond to four Rayleigh channel paths and one LOS path), the system EE for the MA-ICSI scheme rises since the spatial DoFs of five channel paths (more than the number of receive MAs) can be exploited based on the I-CSI. This is different from the schemes based on the S-CSI, where up to four channel paths can be utilized.

\section{Conclusion}
In this paper, we investigated a multi-user MIMO downlink system with S-CSI, where the BS and UTs are equipped with multiple MAs. The system EE for the multi-user scenarios was maximized by optimizing the precoding matrix and the APVs through an AO algorithm. We first resorted to the DE technique to formulate the deterministic minorizing function of the system EE and the deterministic function of each UT's average achievable rate w.r.t. the transmit variables and the associated receive APV, respectively. Subsequently, two SCA algorithms were developed to alternatively optimize the transmit variables and the receive APVs based on the rule of maximizing the above formulated deterministic objective functions, respectively. Finally, the AO algorithm was tailored for the single-user scenario. Numerical results revealed that the proposed MA-enhanced systems can significantly improve the system EE compared to several benchmark schemes utilizing the S-CSI and the optimal performance can be achieved with a finite size of movement regions for MAs. In future work, the power of the RF chains can be optimized by carefully selecting appropriate antenna subsets and turning off the RF chains corresponding to the unselected antennas.

\begin{appendices}
	
\end{appendices}

\end{document}